\documentclass[prd,aps,twocolumn,showpacs]{revtex4}
\usepackage{epsfig}

\newcommand{\bk}{{\mathbf k}}

\newcommand{\bg}{{\mathbf g}}

\newcommand{\by}{{\mathbf y}}
\newcommand{\bx}{{\mathbf x}}
\newcommand{\bz}{{\mathbf z}}
\newcommand{\bn}{{\mathbf n}}
\newcommand{\bv}{{\mathbf v}}
\newcommand{\bnabla}{{\mathbf \nabla}}

\newcommand{\JJ}{{\cal J}}
\newcommand{\SSS}{{\cal S}}

\newcommand{\cd}{\cdot}
\newcommand{\al}{\alpha}
\renewcommand{\b}{\beta}
\newcommand{\de}{\delta}
\newcommand{\De}{\Delta}

\newcommand{\ga}{\gamma}
\newcommand{\Ga}{\Gamma}

\newcommand{\La}{\Lambda}
\newcommand{\la}{\lambda}
\newcommand{\Om}{\Omega}
\newcommand{\om}{\omega}

\newcommand{\dt}{\ensuremath{\delta \theta}}
\newcommand{\dx}{\ensuremath{\delta x}}
\newcommand{\ra}{\rightarrow}

\newcommand{\be}{\begin{equation}}
\newcommand{\ee}{\end{equation}}
\newcommand{\gsim}{\stackrel{>}{\sim}}
\newcommand{\lsim}{\stackrel{<}{\sim}}
\newcommand{\bea}{\begin{eqnarray}}
\newcommand{\eea}{\end{eqnarray}}
\newcommand{\bean}{\begin{eqnarray*}}
\newcommand{\eean}{\end{eqnarray*}}
\newcommand{\dd}{\partial}
\newcommand{\mr}{\mathrm}
\newcommand{\ie}{{\em i.e. }}
\newcommand{\eg}{{\em e.g. }}

\newcommand{\pau}{\ensuremath{^{\alpha\,(1)}}} 
\newcommand{\apt}{{\ensuremath{\mathcal H}}}

\begin{document}

\title{Fluctuations of the luminosity distance}

\author{Camille Bonvin}
\email{camille.bonvin@physics.unige.ch} \affiliation{D\'epartement
de Physique Th\'eorique, Universit\'e de
  Gen\`eve, 24 quai Ernest Ansermet, CH--1211 Gen\`eve 4, Switzerland}

\author{Ruth Durrer}
\email{ruth.durrer@physics.unige.ch} \affiliation{D\'epartement de
Physique Th\'eorique, Universit\'e de
Gen\`eve, 24 quai Ernest Ansermet, CH--1211 Gen\`eve 4, Switzerland}

\author{M.~Alice Gasparini}
\email{alice.gasparini@physics.unige.ch}
\affiliation{D\'epartement de Physique Th\'eorique, Universit\'e
de Gen\`eve, 24 quai Ernest Ansermet, CH--1211 Gen\`eve 4,
Switzerland}

\date{\today}

\begin{abstract}

We derive an expression for the luminosity distance in a
perturbed Friedmann universe. We define the correlation function and
the power spectrum of the luminosity distance fluctuations and express
them in terms of the initial spectrum of the Bardeen potential.
We present semi-analytical results for the case of a pure CDM (cold dark matter)
universe. We argue that the luminosity distance power spectrum
represents a new observational tool which can be used to determine
cosmological parameters. In addition, our results shed some light into
the debate whether second order small scale fluctuations can mimic an
accelerating universe. 

\end{abstract}


\maketitle

\section{Introduction}
\label{sec:intro}

Some years ago, to the biggest surprise for the physics
community, measurements of luminosity distances to far away type
Ia supernovae have indicated that the Universe presently undergoes a
phase of accelerated expansion~\cite{snIa}. If the Universe is
homogeneous and isotropic, \ie a Friedmann-Lema\^\i tre universe,
this means that the energy density is dominated by
some exotic 'dark energy' which obeys an equation of state of the
form $P< -\rho/3$. The best known dark energy candidate is vacuum
energy or, equivalently, a cosmological constant. This discovery
has lately been supported by several other combined data sets,
like the cosmic microwave background (CMB) anisotropies combined with
either large scale structure or measurements of the Hubble
parameter~\cite{WMAP}.

On the other hand, since quite some time, it is known that
locally measured cosmological parameters like $H_0$ or the
deceleration parameter $q_0$ might not be the ones of the
underlying Friedmann universe, but they might be dressed by local
fluctuations~\cite{Buchert}. Therefore, it is of great importance to
derive a general formula of the luminosity distance in a
universe with perturbations. To some extent, this has been done in
several papers before~\cite{Sasaki,Dyer}. But the formula which we
derive here is new. We shall comment on the relations later on. 

Lately, it has even been argued that
second order perturbations might be responsible for the observed
acceleration and that no cosmological constant or dark energy is
needed~\cite{Kolb, Barausse}.
This claim is very surprising, as it seems to
require that back reaction leads to big perturbations out to very
large scales, contrary to 
what is observed in the CMB. This proposal has thus promptly initiated
a heated debate~\cite{2debate}.

On the one hand, the present work is a contribution in this
context. We calculate the measurable luminosity distance in a
perturbed Friedmann universe and determine its fluctuations (within
linear perturbation theory). We show that these remain smaller than one
and therefore higher order perturbations are probably not relevant. 
The main point of our procedure is 
that we use only \textit{measurable} quantities and not
some abstract averaged expansion rate to determine the
deceleration parameter. We actually calculate the luminosity
distance $d_L(\bn, z)$ where $\bn$ defines the direction of the
observed supernova and $z$ its redshift. We then determine the
power spectrum $C_\ell(z,z')$ defined by
\bea \label{eq:defalm}
 d_L(\bn, z) &=& \sum_{\ell m} a_{\ell m}(z)Y_{\ell m}(\bn) \\
C_\ell(z,z') &=& \langle a_{\ell m}(z)a_{\ell m}^*(z') \rangle ~.
\label{eq:defCl}
\eea
Here the $\langle \cdot \rangle$ denotes a statistical average. Like
for the cosmic microwave background, statistical isotropy implies that
the $C_\ell$'s are independent of $m$. 

We then analyze whether the deviations of the angular diameter
distance from its background value can be sufficient to fake
an accelerating universe.

Aside from this problem, the new variable which is defined and calculated
in this paper, might in principle present an interesting and
novel observational tool to determine cosmological parameters.
And this is actually the main point of our work.
We hope to initiate a new observational effort, the measurement of
the luminosity distance power spectrum, with this paper. 
A detailed numerical calculation of the $d_L$ power spectrum
and the implementation of a parameter search algorithm are
postponed to future work. Here we simply show that for large
redshifts, $z\ge 0.4$ and sufficiently high multipoles, $\ell>10$ the
lensing effect dominates. However, at smaller redshift and especially
at low $\ell$'s other terms can become important, most notably the
Doppler term due to the peculiar motion of the supernova.

The paper is organized as follows. In Section~\ref{sec:dL} we
derive a general formula for the luminosity distance valid in
(nearly) arbitrary geometries. In the next section we apply
the formula to a perturbed  Friedmann universe. In
Section~\ref{sec:Cl} we derive general expressions for the $d_L$
power spectrum in terms of the Bardeen potentials.
We then evaluate our expressions in terms of relatively crude
approximations and some numerical calculations for a simple $\Om_M=1$
CDM model in Section~\ref{sec:CDM}.
In Section~\ref{sec:con} we discuss our results and conclude.

\noindent{\bf Notation:} We denote 4-vectors by arbitrary letters,
sometimes with and sometimes without Greek indices,
$k=(k^\mu)$. Three-dimensional vectors are denoted bold face or with
Latin indices, $\by = (y^i)$. We use the metric signature $(-,+,+,+)$.
The covariant derivative of the 4-vector $k$ in direction of the
4-vector $n$ is often denoted by $\nabla_nk \equiv
(n^\mu k^\al{}_{;\mu})$.

\section{The luminosity distance in inhomogeneous geometries}
\label{sec:dL}

\begin{figure}[ht]
\centerline{\epsfig{figure=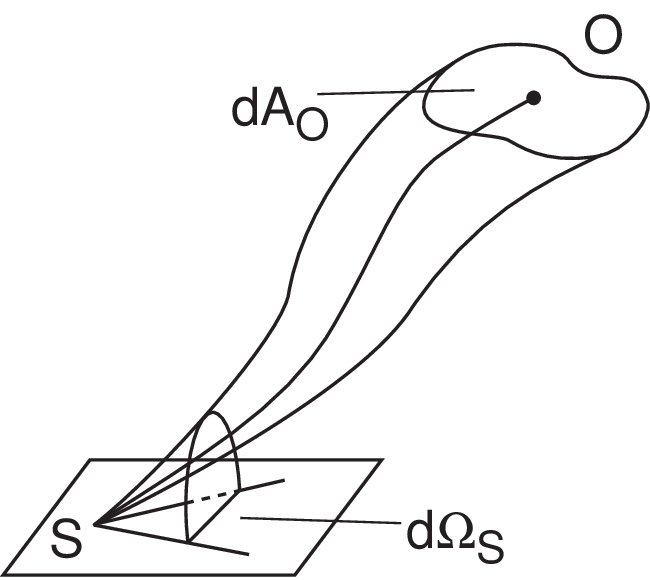,width=5cm}}
\caption{ \label{fig:angle} A light beam emitted at the source event
  S ending on the observer O. At the source position, the plane normal
  to the source four-velocity is indicated.}
\end{figure}

We consider an inhomogeneous and anisotropic universe with
geometry $ds^2 = g_{\mu\nu}dx^\mu dx^\nu$. We place a standard
candle emitting with total luminosity $L$ ( energy per unit proper
time) at spacetime position $S$. Its four-velocity is $u_S$. An
observer at spacetime position $O$ with four velocity $u_O$ (see
Fig.~\ref{fig:angle}) receives the 
energy flux $F$ (energy per unit proper time and per surface). The luminosity
distance between the source at $S$ and the observer at $O$ is
defined by
\be\label{eq:dLdef}
d_L(S,O) = \sqrt{\frac{L}{4\pi F}}~.
\ee
The observer measures the flux $F$ and 'knows' the intrinsic
luminosity $L$ of the standard candle. Furthermore, she determines the
source redshift $z$ 
and direction $\bn$ and thereby obtains the function $d_L(\bn,z)$,
which we now want to express in terms of the spacetime geometry.

Be $d\Om_S$ the infinitesimal solid angle around the source and
$dA(x)$ the  infinitesimal surface element on the surface normal to
the photon beam at the position $x$ along the photon trajectory
from $S$ to $O$, then
\be \label{eq:jacobi}
d_L^2(S,O) = \frac{dA_O}{d\Om_S}(1+z)^2 =\left|\det
J(O,S)\right|(1+z)^2~.
\ee
Here $J$ is the so called Jacobi map mapping initial directions
$\dt^\al_S$  around the source into  vectors $\de x^\mu_O$ transversal to
the photon beam at the observer position~\cite{SEF},
\be\label{eq:jacobi2}
\de x_O^\mu = \JJ^\mu{}_\al(O,S)\dt_S^\al~.
\ee
The factor $1+z = \om_S/\om_O$ is the redshift of the
source. There is a factor $1+z$ due to the redshift of the
emitted energy and a second factor due to the time dilatation in
$F\propto dE_O/d\tau_O$ with respect to $L=dE_S/d\tau_S$. If $k$
denotes the 4-vector of the photon momentum and $u_S$ and $u_O$ are the
source and observer  4-velocities respectively, we have
\be \label{eq:redshiftS}
-\om_S \equiv (k\cdot u_S) = g_{\mu\nu}(S)k^\mu(S)u_S^\nu(S) \quad
\mbox{ and } \ee
\be \label{eq:redshiftO}
-\om_O\equiv (k\cdot u_O) = g_{\mu\nu}(O)k^\mu(O)u_O^\nu(O)
\ee

If we have a standard candle source of which we know $L$ and we
measure $F$, we can therefore determine $\left|\det
J(O,S)\right|^{1/2} \om_S/\om_O$, which contains information about the
spacetime geometry. Of course it also depends on the source and
observer velocities. The Jacobi map $\JJ^\mu{}_\al(O,S)$ maps direction
vectors normal to the photons direction and normal to $u_S$ at $S$ into vectors
normal to the photon direction and 
$u_O$ at $O$. It depends on the source velocity $u_S$ and on the
curvature tensor along the photon geodesic from $S$ to $O$. As we
shall see, it does not depend on the observer velocity $u_O$.

Even though in the form~(\ref{eq:jacobi2}), $J$ is given by the $4\times
4$ matrix $\JJ^\mu{}_\al(O,S)$, we have to take into account that
the vectors $\de x_O^\mu$
as well as $\dt^\al_S$ live in the two dimensional subspace normal to
$u_O$ respectively $u_S$ and normal to the photon direction at $O$  and
$S$. The latter are given by
\bea\label{eq:photondirO}
n_O &=& \frac{1}{\om_O}\left(k(O) + (k(O)\cdot u_O)u_O\right) \quad \mbox{
  and } \\
\label{eq:photondirS}
n_S &=& \frac{1}{\om_S}\left(k(S) + (k(S)\cdot u_S)u_S\right)~.
\eea
The photon direction vectors $n_S$ and $n_O$ are normalized spacelike
vectors pointing into the photon direction in the reference frame of
the source at $S$ and of the observer at $O$ respectively.
Denoting the projectors onto the subspaces normal to $u_S,n_S$ and
$u_O, n_O$ by $P_S$ and $P_O$ we have
\bea\label{eq:proj}
(P_S)^\mu{}_\nu  &=& \de^\mu{}_\nu  + u_S^\mu u_{S\nu} - n_S^\mu n_{S\nu}
\quad \mbox{ and} \\
(P_O)^\mu{}_\nu  &=& \de^\mu{}_\nu  + u_O^\mu u_{O\nu} - n_O^\mu
n_{O\nu} ~.
\eea
The true Jacobi map is $J(O,S) =P_O\JJ P_S$ understood as two dimensional
linear map. For convenience we shall write it as four-dimensional
application and determine its determinant as the product of the
two non-vanishing eigen-values.

To determine the Jacobi map we now derive a differential equation for
the evolution of the difference vector $\de x^\mu(\la)$ in a given direction
$\dt^\al_S$ along the photon trajectory. The final value  $\de
x^\mu(\la_O)$ then depends linearly on the initial conditions
$\dt^\al_S$. For this we denote the photon trajectory by
$f^\al(\la,\bf{0})$ and parameterize neighboring light-like geodesics by
$f^\al(\la,\de\by)$.
The 4-vector
$$ k^\al(\de\by) = \frac{\dd f^\al(\la,\de\by)}{\dd\la} $$
is the tangent of neighboring photons at $\de\by$ and
$$ \de x^\al = \frac{\dd f^\al}{\dd y^i}\de y^i $$
connects the geodesics $f^\al(\la,{\bf 0})$ and $f^\al(\la,\de\by)$.
Since the 'beam' $f^\al(\la,\by)$ describes photons which are all
emitted at the same event $S$ they have the same phase (eikonal)
$\SSS$. With $k_\al = -\nabla_\al\SSS$ we therefore have
\be
0=\nabla_{\de x}\SSS \equiv \de x^\al \nabla_\al\SSS = -\de x^\al k_\al ~.
\ee
In order for the 4-vectors $\de x^\al(\by)$ to sweep a surface normal
to $u_O$ at the
observer event $O$ at $\la=\la_O$, we also need $(\de x(\la_O)\cd u_O)
=0$. This is a priori not true. However, we can
re-parameterize $f$ by
\be
\la \ra \bar\la = \la +h(\by) \quad\mbox{ and } \quad
 \by \ra \bar\by =\bg(\by) ~.
\ee
Under this reparameterization $\de x$ transforms as $\de x^\al \ra
\overline{\de x}^\al =\de x^\al + k^\al \de h$.
It is easy to see that $g_{\al\beta}\de x^\al \de x^\beta =
g_{\al\beta}\overline{\de x}^\al \overline{\de x}^\beta$, hence the
length of the vector $\de x$ is invariant under this
reparameterization. Since $u_O$ is timelike, $(k(\la_O)\cd u_O) \neq 0$ and
we can hence choose a parameterization such that $(\de x(\la_O)\cd u_O)
= 0$.

The directions $\dt^\al$ are given by
\be
 \dt^\al = \frac{1}{\om_S}(\nabla_k\de x)^\al =
 \frac{1}{\om_S}(\nabla_{\de x}k)^\al  ~.
\ee
The last equality requires a brief calculation which can be found, \eg
in~\cite{SEF}. To convince oneself that the above definition of
$\dt^\al$ is suitable, one easily verifies (see \cite{SEF}) that
$\dt^\al_S$ is normal to the source velocity $u_S$ and the photon
direction $n_S$ and that it is normalized.

To find the differential equation for $\de x(\la)$ we use the relations

\begin{eqnarray}
R^\al{}_{\beta\mu\nu}k^\beta &=& \left(\nabla_\mu \nabla_\nu -
\nabla_\nu \nabla_\mu\right)k^\al\nonumber\\
(\nabla_k\de x)^\beta = k^\al\nabla_\al\dx^\beta &=& \dx^\al\nabla_\al k^\beta
= (\nabla_{\de x}k)^\beta ~.
\end{eqnarray}
Furthermore,
\begin{eqnarray}
R^\al_{\beta\mu\nu}k^\beta k^\mu\dx^\nu &=& k^\mu\dx^\nu
\left(\nabla_\mu\nabla_\nu-\nabla_\nu\nabla_\mu\right)k^\al\nonumber\\
&=& \dx^\nu\nabla_k \left(\nabla_\nu k^\al\right)- k^\mu \nabla_{\dx}
\left(\nabla_\mu k^\al\right) \nonumber\\
&=&\nabla_k \left(\dx^\nu\nabla_\nu k^\al\right)- \left(\nabla_k
\dx^\nu \right) \left(\nabla_\nu k^\al\right)  \nonumber\\
&& -\nabla_{\dx}\left(k^\mu\nabla_\mu k^\al\right) + \left(\nabla_{\dx}
k^\mu \right) \left(\nabla_\mu k^\al\right)\nonumber\\
&=& \nabla_k\left(\dx^\nu\nabla_\nu k^\al\right) \nonumber\\
&=&\nabla_k\left(k^\nu\nabla_\nu\dx^\al\right)= \nabla_k(\om_S\dt^\al)~.
\end{eqnarray}
From the third to the fourth line we have used that $\nabla_k\dx =
\nabla_{\dx} k$ and $\nabla_k k =0$.
We therefore obtain the system of equations
\begin{eqnarray}
\nabla_k(\om_S\dt^\al)&=&R^\al_{\b\mu\nu}k^\b k^\mu\dx^\nu \\
\nabla_k(\dx^\al)&=&\om_S\dt^\al ~.
\end{eqnarray}

With the definition of the covariant derivative this finally gives

\begin{eqnarray}
\frac{d(\dx^{\al})}{d\la} &=& -\Gamma^{\al}_{\mu \nu}k^{\mu}\dx^{\nu}
+ \om_S\dt^{\al}  \nonumber\\
&\equiv& C^{\al}_{\nu}(\la)\dx^{\nu} + \om_S\dt^{\al}
 \label{eq:dx}\\
\frac{d(\om_S\dt^{\al})}{d\la} &= &R^{\al}_{\b \mu \nu}k^{\b}k^{\mu}\dx^{\nu}
-\Gamma^{\al}_{\mu \nu}k^{\mu}\om_S\dt^{\nu}  \nonumber\\
&\equiv&
A^{\al}_{\nu}(\la)\dx^{\nu}+C^{\al}_{\nu}(\la)\om_S\dt^{\nu} \label{eq:dth}
\end{eqnarray}
where we have set
\begin{equation} \label{eq:C}
C^{\al}_{\b}(\la)=-\Gamma^{\al}_{\mu \b}k^{\mu} \quad \textrm{and}
\quad A^{\al}_{\b}(\la) = R^{\al}_{\rho \mu \b}k^{\rho}k^{\mu} ~.
\end{equation}
We now define
\begin{equation}
\vec{Z}=\left( \begin{array}{c}
        \dx^{\al}\\
        \om_S\dt^{\al}
        \end{array} \right) ~.
\end{equation}
This (8 component) vector then satisfies the equation

\begin{equation}
\frac{d\vec{Z}(\la)}{d\la}=B(\la)\vec{Z}(\la)
\label{eq:syseq}
\end{equation}
with
\begin{equation}
B(\la)=\left( \begin{array}{cc}
            C^{\al}_{\b}(\la) & \delta^{\al}_{\b} \\
            A^{\al}_{\b}(\la) & C^{\al}_{\b}(\la)
            \end{array} \right) ~.
\end{equation}
The initial conditions are $\dx^{\al}(\la_S)=0$ since all photons
start from the same source event and  $\left(k^{\al}\dt_{\al}\right)(\la_S) =
\left(u^\al_S\,\dt_\al(\la_S)\right) =0$ as
we have seen above. The solution of Eq.~(\ref{eq:syseq}) therefore
provides a linear relation between the initial condition
$\dt^{\al}(\la_S)$ and $\dx^\al(\la)$,
\be
\dx^{\al}(\la)=\JJ^{\al}_{\b}(\la)\dt^{\b}(\la_S) ~.
\ee
With $\JJ(\la_O)$ we can then easily determine the true Jacobi map
$J(O,S) = P_O\JJ(\la_O) P_S$.

\section{The luminosity distance in a perturbed Friedmann universe}

\subsection{Conformally related luminosity distances}
We consider two geometries related by
\begin{equation}
d\tilde s^2 =\tilde g_{\mu\nu}dx^\mu dx^\nu = a^2(x)g_{\mu\nu}dx^\mu
dx^\nu = a^2(x)ds^2~.
\end{equation}
We want to relate the angular diameter distances of the two metrics.
If $\tilde k$ is a light-like geodesic for the metric $d\tilde s^2$ with
affine parameter $\tilde\la$, then $k = a^2\tilde k$ is a light-like
geodesic for $ds^2$ with affine parameter $\la$ determined by
\[
\frac{d\tilde\la}{d\la} = a^2~.
\]
Furthermore, be $\tilde u^\mu = \frac{dx^\mu}{d\tilde \tau}$ the
4-velocity of an observer with metric  $d\tilde s^2$ and be
$\tilde\tau$ its proper time such that $\tilde g_{\mu\nu}\tilde u^\mu
\tilde u^\nu  =- 1$, then $ u^\mu = \frac{dx^\mu}{d\tau}$ is the
corresponding 4-vector of the observer with respect to the metric
$ds^2$ with proper time $\tau$ if $\frac{d\tilde \tau}{d\tau} = a$. In
other words
\be
\tilde u^\mu = \frac{dx^\mu}{d\tilde \tau} =\frac{dx^\mu}{d\tau}
\frac{d\tau}{d\tilde\tau} = a^{-1}u^\mu~.
\ee
The redshift of a photon emitted at $S$ and observed at $O$
determined with respect to the two metrics is therefore related by
\be
1 +\tilde z = \frac{\tilde\om_S}{\tilde\om_O} = \frac{(\tilde
  g_{\mu\nu} \tilde k^\mu\tilde u^\nu)_S}{(\tilde
  g_{\mu\nu} \tilde k^\mu\tilde u^\nu)_O} = \frac{a_O(g_{\mu\nu} k^\mu
  u^\nu)_S}{a_S( g_{\mu\nu} k^\mu u^\nu)_O} = \frac{a_O}{a_S}(1+z) ~.
\ee

To determine the relation between the Jacobi maps $J^\al{}_\beta
=\frac{\dx_O^\al}{\de\theta_S^\beta}$ we just have to remember that
angles are not affected by conformal transformations, but distances
scale with the conformal factor $a$. Therefore
\bea
\tilde J(S,O) &=& \frac{\de\tilde x_O^\al}{\de\theta_S^\beta}=  a_O
\frac{\dx_O^\al}{\de\theta_S^\beta}= a_O J(S,O)~, \\
 \det\tilde J(S,O) &=& a_O^{2} \det J(S,O)~.
\eea
For the angular distance relation we finally obtain
\bea
\tilde d_L &=& (1+\tilde z)\sqrt{|\det\tilde J(S,O)|}  \nonumber \\
           & =& \frac{a_O^2}{a_S}(1+z)\sqrt{|\det J(S,O)|} =
\frac{a_O^2}{a_S}d_L ~. \label{eq:dist.conf}
\eea

This relation is very useful in Friedmann cosmology. The Friedmann
metric is given by
\be\label{eq:Frmetric}
d\tilde s^2 = a^2\left(-d\eta^2 +\ga_{ij}dx^idx^j\right) =a^2ds^2
\ee
where $\ga$ is the metric of a 3--space with constant curvature $K$. The
luminosity distance of a photon emitted at conformal time $\eta_S$ and
observed at $\eta_O$ with respect to the metric $ds^2$ is simply
$\eta_O-\eta_S = \int_{\eta_S}^{\eta_O}d\eta$. The Friedmann equation
for a universe containing matter, radiation, curvature and a
cosmological constant reads
\be\label{eq:Fried}
\left(\frac{\dot a}{a}\right)^2 = H_0^2\left[\Om_ma^{-1}
  +\Om_\mr{rad}a^{-2} +\Om_K +\Om_\La a^2 \right] ~,
\ee
where we have normalized $a_O =1$ and we have introduced the density
parameters $\Om_{m} = \rho_{m}(\eta_O)/\rho_c(\eta_O)$, $\Om_
\mr{rad} = \rho_\mr{rad}(\eta_O)/\rho_c(\eta_O)$, 
$\Om_K = -K/H_0^2$ and $\Om_\La = \La/(3H_0^2)$.

After the variable transformation to $z+1 =1/a$,  $dz =-da/a^2$ we obtain
\[
 d\eta = \frac{ H_0^{-1} ~ dz}{\sqrt{\Om_\mr{rad}(1+z)^4 +\Om_m(z+1)^3
  +\Om_K(z+1)^2 +\Om_\La}}~.
\]
This leads to the well known expression
for the luminosity distance to an object emitting at redshift $z_S$
observed today at $z_O =0$,
\bea
d_L(z_S)^{\mr{Friedman}} ~  = ~ \frac{\eta_0-\eta_S}{a_S} ~ = \quad
  \qquad \qquad\qquad  && \nonumber \\
 \frac{1+z_S}{H_0}\int_1^{z_S+1}\!\! \frac{dx}{ \sqrt{\Om_\mr{rad}x^4 +
     \Om_mx^3 +\Om_Kx^2 +\Om_\La}} ~.  \label{eq:lumFried}
\eea
Comparing this expression with the measured luminosity distance from
 supernovae type Ia at different redshifts has led to the claim that
 the cosmological constant be  non-vanishing~\cite{snIa}.

\subsection{The Jacobi map in a perturbed  Friedmann universe}
We now consider a  Friedmann universe with scalar perturbations. In
longitudinal (or Newtonian) gauge the metric is given by
\be
\tilde g_{\mu\nu}dx^\mu dx^\nu = a^2\left[ -(1+2\Psi)d\eta^2
+(1-2\Phi)\ga_{ij}dx^idx^j\right] ~.
\ee
For perfect fluids the metric perturbations $\Psi$ and $\Phi$ are
equal. We assume in the sequel $\Phi=\Psi$. Furthermore, we consider a
spatially flat universe ($K=0$), so that $\ga_{ij}=\de_{ij}$.

We now determine the luminosity distance for the metric
\be\label{eq:conf}
ds^2 =  -(1+2\Psi)d\eta^2 +(1-2\Psi)\de_{ij}dx^idx^j ~.
\ee
We then relate this to the physical luminosity distance via
the relation (\ref{eq:dist.conf}).

We assume that the galaxy containing the supernova as well as the one
containing the observer are moving with the cosmic fluid. To first
order in the perturbations, the four velocity of the cosmic fluid is
given by
\be
(u^\mu) = (1-\Psi, v^i) ~,
\ee
where $v^i$ is the peculiar velocity field.

\subsubsection{Redshift}

The photon geodesic is obtained by integrating the geodesic equation
to first order. Since the background is Minkowski,
the background photon momentum is constant and we may normalize the
affine parameter such that
$\bar k^0 =1$ and $\bar k^i =n^i$ with $\sum_1^3 n^in^i =1$.
Here over-bars denote background quantities. For the perturbed
4-velocity of the photon we may still assume $ k^0_S =1$.
The geodesic equation then gives (to first order)
\bea
k^0(\la_O)- k^0(\la_S) &=& k^0(\la_O) - 1 =
-2\int_{\la_S}^{\la_O}d\la\nabla\Psi(\la)\cdot\bn  \nonumber \\
 &=& -\left. 2\Psi\right|^O_S + 2\int_{\la_S}^{\la_O}d\la\dot\Psi
 \qquad \mbox{ and}  \nonumber\\
k^i(\la_O)- k^i(\la_S)&=& 2n^i(\Psi_O -\Psi_S)
-2\int_{\la_S}^{\la_O}d\la \dd_i\Psi(\la)  ~. \nonumber
\eea
The redshift of a photon emitted at spacetime position $S$ and
observed at $O$  then becomes

\be
1 +z = \frac{(g_{\mu\nu}k^\mu u^\nu)_S}{(g_{\mu\nu}k^\mu u^\nu)_O} = 1
+ \left[\Psi +\bv\cdot \bn \right]^O_S -2\int_{\la_S}^{\la_O}d\la\dot\Psi ~.
\ee

\subsubsection{The perturbed Jacobi map}
To determine the Jacobi map we have to solve the system
(\ref{eq:syseq}) to first order. We first determine the maps
$C^\al{}_\beta$ and $A^\al{}_\beta$ which make up the matrix $B^N{}_M$.
According to Eq.~(\ref{eq:C}), $C^\al{}_\beta =
-\Ga^\al_{\beta\ga}k^\ga$. Since $\Ga^\al_{\beta\ga}$ is already first
order, we may insert the zeroth order expression for $k^\ga$ leading
to
\bea
C^0_0 &=& -\Psi' \nonumber\\
C^0_i &=& - \dd_i\Psi+\dot{\Psi}n^i\nonumber\\
C^i_0 &=& -\dd_i\Psi +\dot{\Psi}n^i  \nonumber\\
C^i_j &=& \Psi'\delta^i_j + \dd_j\Psi  n^i -\dd_i\Psi n^j ~.
\eea
Here we denote the derivative along the geodesic with a prime and the
derivative w.r.t. conformal time by an over-dot, $\frac{d}{d\la} \equiv
 {}'$ and  $\frac{\dd}{\dd\eta} \equiv \dot{ }$.
The matrix $A$ is given by
$A^{\al}_{\beta}=R^{\al}_{\mu\nu\beta}k^{\mu}k^{\nu}$. Again, since
$R^{\al}_{\mu\nu\beta}$ is of first order, we may insert the zeroth
order expression for the photon velocity. Note that $A_{\al\beta}$,
  unlike $C_{\al\beta}$, is symmetric.  Computing the Riemann
tensor of our perturbed metric we obtain
\begin{eqnarray}
A^0_0 &=& 2\ddot{\Psi} +\frac{d^2\Psi}{d\la^2}-2\frac{d\dot{\Psi}}{d\la}
\nonumber\\
A^0_i&=&2\dd_i\dot{\Psi} -\frac{d\dd_i\Psi}{d\la} -
  \frac{d\dot\Psi}{d\la}n^i\nonumber\\
A^i_0 &=& - A^0_i\nonumber\\
A^i_j&=&-\frac{d^2\Psi}{d\la^2}\delta^i_j-2\dd_j\dd_i\Psi
+\frac{d\dd_j\Psi}{d\la}n^i + \frac{d\dd_i\Psi}{d\la} n^j ~.\nonumber\\
\end{eqnarray}

The Christoffel symbols and the Ricci tensor of the perturbed metric
are given in Appendix~\ref{ap:geo}. Spatial indices $i$ or $j$ are
raised and lowered with the flat metric $\de_{ij}$. Therefore, no
special attention is paid to their position.

To solve it, we now split the system (\ref{eq:syseq}) into its zeroth
and first order components,

\begin{equation}
\vec{Z}=\vec{Z}^{(0)}+\vec{Z}^{(1)} \quad \mbox{ and } \quad
B = \bar B + B^{(1)}
\end{equation}
To zeroth order, the photons move along straight lines and the energy
is not redshifted so that we simply obtain $\de\bar\theta^\al(\la) =
\de\theta^\al_S$, $\bar\om(\la) =\om_S$ and $\de\bar x^\al(\la)
=(\la-\la_S)\om_S\de\bar\theta^\al_S$. For the Jacobi map this
implies $ \bar\JJ^\al_\beta = (\la_O-\la_S)\om_S\de^\al_\beta$. The
projector onto the tangent space normal to the observer velocity and
the photon direction is simply $\bar P_S =\bar P_O =\bar P$, where
\begin{eqnarray}
\bar P^0_0 &=& \bar P^0_i =\bar P^i_0 = 0\nonumber\\
\bar P^i_j &= &\delta^i_j- n^i n_j~.
\end{eqnarray}
The zeroth order 2-dimensional Jacobi map is therefore given by
$\bar J^\al_\beta =(\bar P\bar\JJ\bar P)^\al_\beta$
\begin{eqnarray}
\bar J^0_0 &=& \bar J^0_i =\bar J^i_0 = 0\nonumber\\
\bar J^i_j &= & (\la_O-\la_S)\om_S\left(\delta^i_j- n^i n_j\right)~.
\end{eqnarray}
The 2-dimensional determinant of the Jacobi map is therefore $\det\bar J =
(\la_O-\la_S)^2 \om_S^2$, leading to the flat space luminosity
distance $d_L =\la_O-\la_S =\eta_O-\eta_S$. For the last equality we
have used that $\bar n^0 = \frac{d\eta}{d\la} =1$. In an unperturbed
Friedmann universe this reproduces (\ref{eq:lumFried}).

Since $C$ and $A$ are already first order, the first order
differential equation becomes
\onecolumngrid

\begin{eqnarray}
\frac{d}{d\la}\dx\pau(\la) & =& C\pau_\beta(\la)\de\bar x^\beta(\la) +
(\om_S\dt^\al)^{(1)}(\la)\nonumber\\
\frac{d}{d\la}(\om_S\dt^\al)^{(1)}(\la) &=& A\pau_\beta(\la)
\de\bar x^\beta(\la)+
C\pau_\beta(\la)\bar\om_S \delta\bar\theta^\beta(\la)~.\nonumber\\
\end{eqnarray}
Making use of the background solution we obtain
\begin{eqnarray}
(\dt^\al)^{(1)}(\la) &=& \int_{\la_1}^{\la}d\la'
  \left(A^\al_\beta(\la') (\la'-\la_S)+C^\al_\beta(\la')\right)
 \de\bar\theta^\beta_S +(\dt^\al_S)^{(1)}  \\
\dx\pau(\la)&=&\left[\int_{\la_S}^{\la}d\la'C^\al_\beta(\la')(\la'-\la_S)+
\int_{\la_S}^{\la}d\la' \int_{\la_S}^{\la'} d\la''
  \left(A^\al_\beta(\la'')(\la''-\la_S) + C^\al_\beta(\la'')\right)
  \right]\bar\om_S\de\bar\theta^\beta_S   \nonumber\\   &&
\quad + (\la-\la_S)(\om_S\dt^\al_S)^{(1)}~.
\end{eqnarray}
The first order contribution to the unprojected Jacobi map then becomes
\bea
\om_S^{-1}\JJ\pau_\beta(\la_O) &=& \int_{\la_S}^{\la_O}d\la
C^\al_\beta(\la) (\la-\la_S)+
\int_{\la_S}^{\la_O}d\la \int_{\la_S}^{\la}d\la'\left(A^\al_\beta(\la')
(\la'-\la_S)+C^\al_\beta(\la')\right) ~.
\eea
We want to calculate

\begin{equation}
J^{(1)}= (P_O\JJ P_S)^{(1)}= \bar P_O\JJ^{(1)} \bar P_S +
P^{(1)}_O \bar \JJ \bar P_S + \bar P_O\bar\JJ P^{(1)}_S ~.
\end{equation}
A short calculation, inserting our results for $C$ and $A$ gives
\bea
&& \left(\bar P_O\JJ^{(1)}\bar P_S\right)^i{}_j  = U\cdot\left(\de^i_j
-n^in_j\right) +W^i_j
-n^in^kW_{kj} -n_jn^kW^i_k +n^in_j n^kn^lW_{kl}
\eea
with

\begin{equation}
U = -2\Psi_S(\la_O-\la_S) + 2\int_{\la_S}^{\la_O}d\la\Psi(\la)
\quad\mbox{ and }\quad 
W_{ij} = -2\int_{\la_S}^{\la_O}d\la
\int_{\la_S}^{\la}d\la'\dd_i\dd_j\Psi(\la')(\la'-\la_S) ~.
\end{equation}
Implicit summation over repeated (spatial) indices is assumed and $n^i
=n_i$, $W^i_j = W_{ij} = W^{ij} $.

Calculating also the first order contributions to the projections we
finally obtain

\begin{eqnarray}
J^0_{\;0} &=& 0 \nonumber\\
J^0_{\;i} &=& \om_S(\la_O-\la_S)\left(v^i_O- n^in_kv^k_O\right) \nonumber\\
J^i_{\;0} &=& \om_S (\la_O-\la_S)\left(-v^i_S+n^in^kv^k_S\right) \nonumber\\
J^i_{\;j} &=& \om_S(\la_O-\la_S) \Bigg\{\left(1-2\Psi_S +
\frac{2}{\la_O-\la_S}\int_{\la_S}^{\la_O}d\la \Psi(\la)\right)
\delta^i_j
 + n^in_j\Bigg(-1+ 2\Psi_S - \frac{2}{\la_O-\la_S}
\int_{\la_S}^{\la_0} d\la\Psi(\la)  \nonumber\\ && 
- \bn(\bv_O +\bv_S)
-2\int_{\la_S}^{\la_0}d\la\mathbf{\nabla}\Psi(\la)\mathbf{n} + 2\mathbf{n}\cdot\mathbf{k}^{(1)}_S\Bigg) +
n^i v_{O~j}+n_jv^i_S 
+2\int_{\la_S}^{\la_0}d\la\dd_j\Psi(\la)n^i -n^ik^{(1)}_{Sj}-n_jk^{(1)i}_S
 \nonumber\\ &&   -\frac{2}{\la_0-\la_S}
\int_{\la_S}^{\la_0}d\la\int_{\la_S}^{\la}d\la'(\la'-\la_S)
\Bigg(\dd_i\dd_j\Psi
-n^in^k\dd_j\dd_k\Psi
-n^jn^k\dd_i\dd_k\Psi
+ n^in^jn^kn^l\dd_k\dd_l\Psi\Bigg)(\la')\Bigg\} ~.
\end{eqnarray}

Like in the unperturbed case, the two eigenvalues of the Jacobi map
are equal. This is due to the fact that the shear contribution to the
Jacobi map still vanishes in first order. A short computation gives the
eigenvalues $\al$, 

\begin{eqnarray}
\al &=& \om_S(\la_O-\la_S)\Bigg\{1 -2\Psi_S +
\frac{2}{\la_O-\la_S}\int_{\la_S}^{\la_O}d\la\Psi(\la)
 \nonumber\\  &&   \qquad \qquad \qquad \qquad
-\frac{1}{\la_O-\la_S} \int_{\la_S}^{\la_O}d\la
\int_{\la_S}^{\la}d\la' (\la'-\la_S)\Big(\nabla^2\Psi(\la')
-\dd_{i}\dd_j \Psi(\la')n^in^j\Big)\Bigg\} ~.
\label{eq:valpropre}
\end{eqnarray}

The luminosity distance of the perturbed Minkowski spacetime it
 given by $d_L = (\om_S/\om_O)\al$. Inserting the above expressions
and taking into account the perturbation of the emission frequency,
$\om_S =-(g_{\mu\nu}k^\mu u^\nu)_S = \bar\om_S +\om_S^{(1)}$, we obtain

\begin{eqnarray}
d_L&=&(\eta_O-\eta_S)\Bigg\{1-2\Psi_O+\Psi_S + \bn\cdot(\bv_O -2\bv_S) +
\frac{2}{\eta_O-\eta_S}\int_{\eta_S}^{\eta_O}d\eta \Psi +
2\int_{\eta_S}^{\eta_0}d\eta\mathbf{\nabla}\Psi \cd\bn
 \nonumber\\  &&
+\frac{2}{\eta_O-\eta_S} \int_{\eta_S}^{\eta_0}d\eta
\int_{\eta_S}^{\eta}d\eta' \mathbf{\nabla}\Psi\cd\bn
 - \frac{1}{\eta_O-\eta_S} \int_{\eta_S}^{\eta_0}d\eta
\int_{\eta_S}^{\eta}d\eta'(\eta'-\eta_S)
\Big(\nabla^2\Psi
-n^in^j\dd_{i}\dd_j \Psi\Big) \Bigg\}\, .
\end{eqnarray}
Here we have also transformed the parameter $\la$ into the conformal
time $\eta$ via the relation 
\[ \frac{d\eta}{d\la} = n^0(\la) = 1 -
2\int_{\la_S}^{\la}d\la'\mathbf{\nabla}\Psi\cdot\bn~.  \]
Now $\eta$ is parametrizing the unperturbed photon geodesic and we
interpret the potential as a function of $\eta$, $\Psi(\eta) =
\Psi(\eta,\bx(\eta))$. We use the notation $\dot\Psi
\equiv\dd_\eta\Psi$, so that
$\frac{d\Psi}{d\eta} =\dot\Psi+\bn\cd\bnabla\Psi$.  
We now also take into account expansion, which gives $\tilde d_L =
\frac{a_O^2}{a_S}d_L$.

Furthermore, we relate the peculiar velocities to the Bardeen
potential via the first oder perturbations of
Einstein's equations. Setting $(\tilde u^\mu) = a^{-1}(1-\Psi, v^i)$ gives~\cite{D94},
\begin{equation}
 v^i(\eta)=-\frac{1}{4\pi Ga^2(\rho +p)} \left(\frac{\dot{a}}{a}
\dd_i\Psi + \dd_i\dot\Psi\right) ~.
\end{equation}

With this we find the following result for the luminosity distance in
an perturbed  Friedmann universe

\begin{eqnarray}
\tilde{d}_L(\eta_S,\bn)&=&\frac{a_O^2}{a_S}(\eta_O-\eta_S)\Bigg\{1-2\Psi_O +\Psi_S+
\mathbf{v}_O\cd\bn
+ \frac{2}{\eta_O-\eta_S}\int_{\eta_S}^{\eta_O}d\eta \Psi
 + 2\bn\cd\Bigg[ \int_{\eta_S}^{\eta_O}d\eta\mathbf{\nabla}\Psi
\nonumber\\ &&  \qquad
+\frac{1}{\eta_O-\eta_S}\int_{\eta_S}^{\eta_O}d\eta
\int_{\eta_S}^{\eta}d\eta'\mathbf{\nabla}\Psi
+\frac{1}{4\pi Ga_S^2(\rho+p)(\eta_S)}
\left(\apt\bnabla\Psi+\bnabla\dot{\Psi}\right)(\eta_S)
\Bigg]
\nonumber\\ &&  \qquad
- \frac{1}{\eta_O-\eta_S} \int_{\eta_S}^{\eta_O}d\eta
\int_{\eta_S}^{\eta}d\eta' (\eta'-\eta_S)\Big(\nabla^2\Psi
- n^in^j\dd_i\dd_{j}\Psi \Big) \Bigg\} ~,  \label{eq:dL(etaS)}
\end{eqnarray}
where we have introduced $\apt \equiv \dot a/a = a^{-1}\frac{da}{d\eta}
\equiv Ha$.  In what follows, we further simplify the formulas by
normalizing the scale factor to 
$ a_O\equiv 1~.$

Here we have used the linear perturbation theory solution for
the source velocity $v_S$. One might argue that the supernovae are
highly non-linear objects inside galaxies and do not move with the
velocity obtained from linear perturbation theory. However, we shall
be interested in distances and angles which are sufficiently large so that the
non-linear contributions to the supernova velocities are uncorrelated
and therefore considering only the linear part of it in the correlation
function is sufficient.

Eq.~(\ref{eq:dL(etaS)})
 is the luminosity distance of a source in direction $-\bn$ 
at conformal time $\eta_S$. However, this quantity
is not directly measurable. What we do measure instead is the redshift
of the source, 
$z_S =\bar z_S +\de z_S$, where $\bar z_S +1 = 1/a(\eta_S)$. Now
\be
 \tilde{d}_L(\eta_S,\bn) =  \tilde{d}_L(\eta(\bar z_S),\bn) \equiv 
 \tilde{d}_L(\bar z_S,\bn) =  \tilde{d}_L(z_S,\bn) - \frac{d}{d\bar z_S}
 \tilde{d}_L(z_S,\bn)\de z_S ~.
\ee
Furthermore,
\bea
  \frac{d}{d\bar z_S} \tilde{d}_L( z_S,\bn) &=& (1+z_S)^{-1} \tilde
  d_L + \apt^{-1}_S  + \mbox{ first order} \qquad \mbox{and } \nonumber \\
\de \tilde z_S &=& (1+z_S)\de z_S = (1+z_S)\left[\Psi_S - \Psi_O
+2\int_{\eta_S}^{\eta_O}d\eta\bn\cd\bnabla\Psi + (\bv_O-\bv_S)\cd\bn\right]~. 
\eea
Inserting this in  Eq.~(\ref{eq:dL(etaS)}) leads to

\begin{eqnarray}
\tilde{d}_L(z_S,\bn)&=&(1+z_S)\Bigg\{(\eta_O-\eta_S)
-\frac{1}{\apt_S}\left(\Psi_S + \bv_O\cd\bn\right)
- \left(\eta_O-\eta_S -  \apt_S^{-1}\right)\Psi_O
 + 2\int_{\eta_S}^{\eta_O}d\eta \Psi
\nonumber\\ &&  \quad
 + 2\bn\cd\Bigg[ -\frac{1}{\apt_S}
\int_{\eta_S}^{\eta_O}\! \hspace{-2mm} d\eta\bnabla\Psi
+\int_{\eta_S}^{\eta_O}\! \hspace{-2mm} d\eta
\int_{\eta_S}^{\eta}\!\!  d\eta'\bnabla \Psi
+ \frac{\eta_O-\eta_S - \apt_S^{-1}}{8\pi Ga_S^2(\rho+p)(\eta_S)}
\left(\apt\bnabla\Psi+\bnabla\dot{\Psi}\right)(\eta_S)
\Bigg]
\nonumber\\ &&  \qquad
- \int_{\eta_S}^{\eta_O}d\eta
\int_{\eta_S}^{\eta}d\eta' (\eta'-\eta_S)\Big(\nabla^2\Psi
-\dd_i\dd_{j}\Psi n^in^j \Big) \Bigg\} ~.  \label{eq:dL(zS)}
\end{eqnarray}
After several integrations by part, one can also derive the following
expression for the luminosity distance, which can also be found
elsewhere~\cite{Sasaki,Barausse}, where it has been derived using the evolution
equations of the expansion and the shear.
\begin{eqnarray}
\tilde{d}_L(z_S,\bn)&=&(1+z_S)(\eta_O-\eta_S)\Bigg\{1
-\frac{1}{(\eta_O-\eta_S)\apt_S}\bv_O\cd\bn
-\left(1 -  \frac{1}{(\eta_O-\eta_S)\apt_S}\right)\bv_S\cdot\bn
\nonumber\\ && 
- \left(1- \frac{1}{(\eta_O-\eta_S)\apt_S}\right)\Psi_S  
-\frac{1}{(\eta_O-\eta_S)\apt_S}\Psi_O
\nonumber\\ && \hspace{-0.5cm}
 + \frac{2}{(\eta_O-\eta_S)}\int_{\eta_S}^{\eta_O}d\eta \Psi
 + \frac{2}{(\eta_O-\eta_S)\apt_S}\int_{\eta_S}^{\eta_O}d\eta \dot\Psi
-2\int_{\eta_S}^{\eta_O}d\eta
\frac{(\eta-\eta_S)}{(\eta_O-\eta_S)}\dot\Psi
+\int_{\eta_S}^{\eta_O}d\eta
\frac{(\eta-\eta_S)(\eta_O-\eta)}{(\eta_O-\eta_S)}\ddot\Psi
\nonumber\\ &&  \quad
-\int_{\eta_S}^{\eta_O}d\eta
\frac{(\eta-\eta_S)(\eta_O-\eta)}{(\eta_O-\eta_S)}\nabla^2\Psi \Bigg\}
 ~.  \label{eq:dL(zS)2}
\end{eqnarray}
\twocolumngrid
A detailed derivation of this result starting from
Eq.~(\ref{eq:dL(zS)}) is given in Appendix~\ref{ap:lit}.
In this equation the first line, apart from the background
contribution, contains the terms  due to peculiar motion of the observer and
emitter (Doppler terms). The second line can be identified as
'gravitational redshift'. This is, however, not entirely correct since
this term does not vanish even if $\Psi_S=\Psi_O$. The third line
collects integrated effects proportional to line of sight integrals of
$\Psi$ and its time derivative, and the fourth and last line represents
the lensing term with $\nabla^2\Psi \propto \de\rho$. This term has
been discussed in the literature before~\cite{lens}. 
An equivalent of the above formula can also be found in~\cite{Lam}.

Eqs.~(\ref{eq:dL(zS)}) and~(\ref{eq:dL(zS)2}) are the final
expressions for the luminosity distance in a perturbed 
Friedmann universe, as a function of the measured source redshift
$z_S$ and its direction $-\bn$.
In the next section we determine the luminosity distance power spectrum
which is, in principle, an observable quantity. 

\section{The luminosity distance power spectrum}\label{sec:PS}
\label{sec:Cl}

We now want to determine the power spectrum of the perturbed
luminosity distance, as defined in the
introduction. For notational simplicity, we drop the $\;\tilde{ }\;$
and use $d_L$ to denote the luminosity distance in a perturbed Friedman
universe. From Eqs.~(\ref{eq:defalm}) and (\ref{eq:defCl}) and the
addition theorem for spherical harmonics, one obtains the correlation function
\bea
\bar d_L(z_S)^{-1}\bar d_L(z_{S'})^{-1} \langle
d_L(z_S,\bn)d_L(z_{S'}\bn') \rangle = \qquad && \nonumber\\
\label{eq:Clcorr} 
\sum_\ell 
\frac{2\ell +1}{4\pi} C_\ell(z_S,z_{S'})P_\ell(\bn\cd\bn') &&
\eea
where $P_\ell$ is the Legendre polynomial of order $\ell$.

\subsection{The dipole}

\begin{figure}[ht]
\centerline{\epsfig{figure=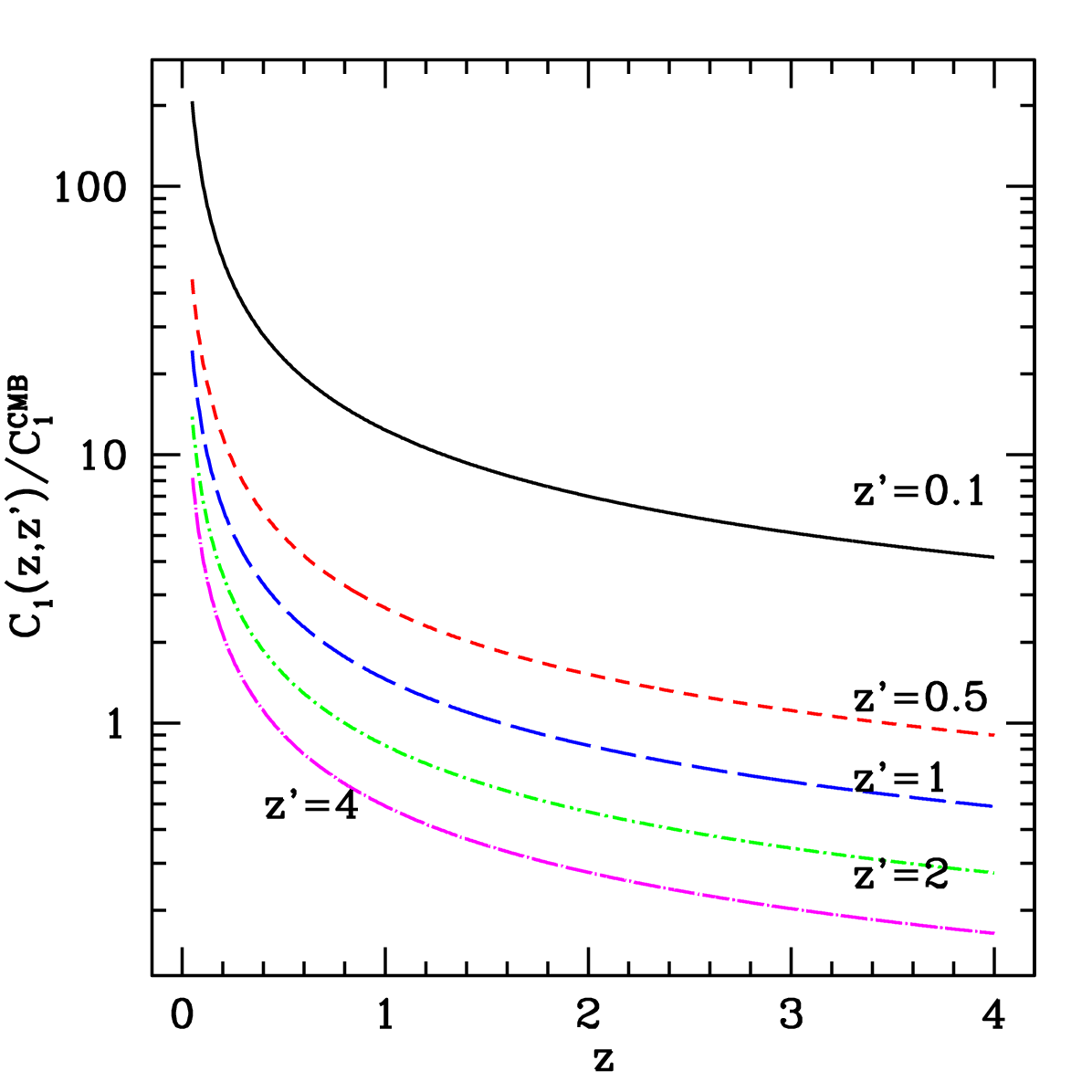,width=6.5cm}}
\caption{ \label{fig:dip1} We show the dipole amplitude in a pure CDM
  universe in units of
  the CMB dipole as a function of $z$ for $z'=0.1$, $0.5$, $1$, $2$
  and $4$ from top to bottom.
}
\end{figure}
\begin{figure}[ht]
\centerline{\epsfig{figure=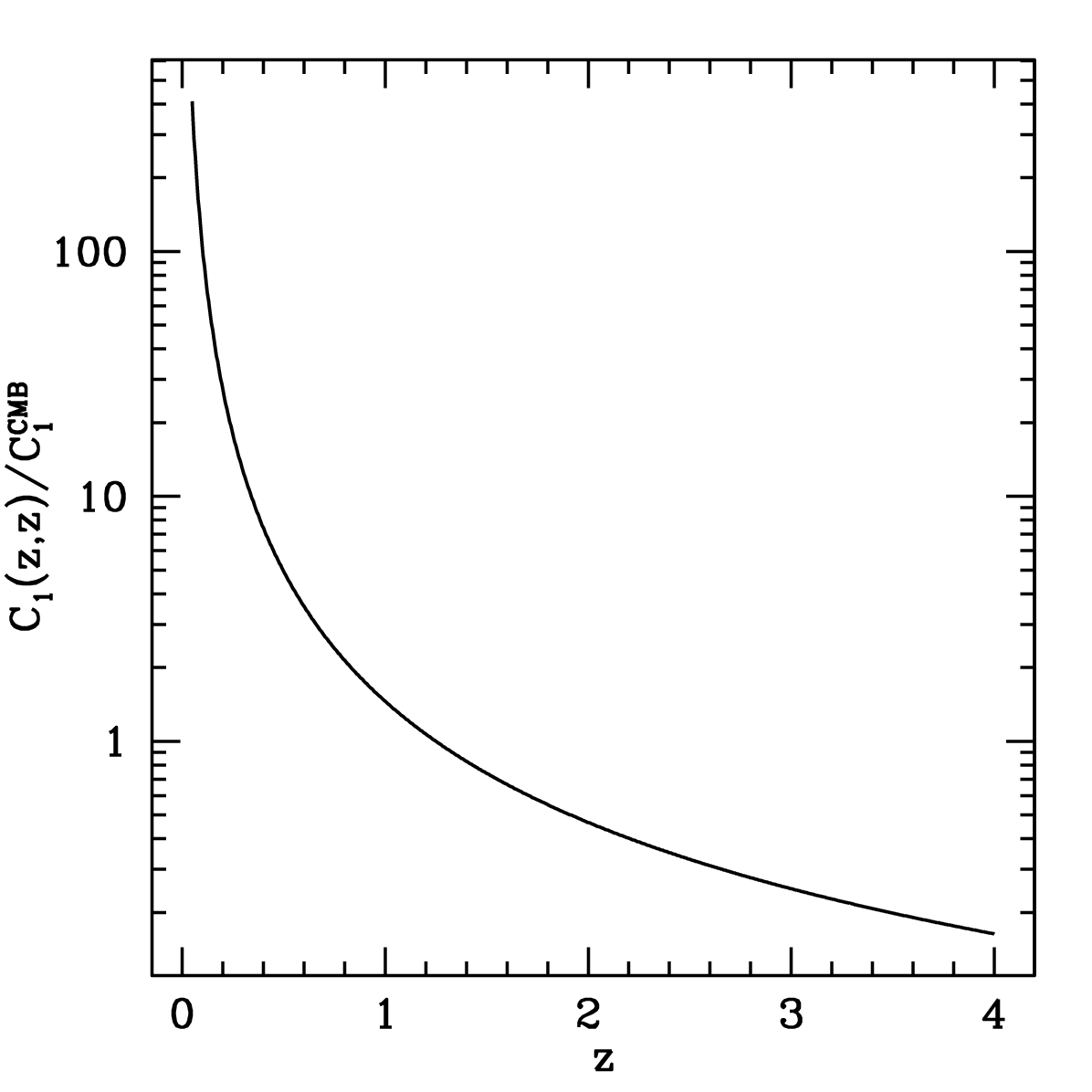,width=6.5cm}}
\caption{ \label{fig:dip2} We show the dipole amplitude in units of
  the CMB dipole as a function of $z=z'$ in a pure CDM universe.
}
\end{figure}

Let us first briefly look at the dipole coming from the peculiar
motion of the observer, the term containing the scalar product 
$\bn\cd \bv_O$. The power spectrum of this term is given by
\bea
\langle d_L^{(v)}(z_S,\bn) d_L^{(v)}(z_{S'},\bn')\rangle = \qquad\qquad &&
\nonumber \\  \qquad 
 \frac{(z_S+1)(z_{S'} +1)}{3\apt_S\apt_{S'}}\langle v_O^2\rangle
   (\bn\cd\bn') ~. &&
\eea
We assume that, like for the anisotropies in the cosmic microwave
background, this term completely dominates the dipole. 
The luminosity distance dipole therefore has the same
direction as the CMB dipole. To determine its amplitude we insert
 $\bar d_L(\eta_S) =(z_S+1)(\eta_O-\eta_S)$. We then obtain
\be
C_1 = \left[\frac{4\pi}{9}\langle v_O^2\rangle\right]
\frac{\apt_S^{-1}\apt_{S'}^{-1}}{(\eta_O-\eta_S)(\eta_O-\eta_{S'})} ~.
\ee
The CMB dipole is given by the
expression in square brackets. In a pure CDM universe with
$\apt=2/\eta$ and $\eta_O/\eta_S = \sqrt{z_S+1}$ we
obtain for the amplitude of the  luminosity distance dipole 
\be
C_1(z,z') = C_1^{\mr{CMB}} 
   \frac{1}{4(\sqrt{z+1} -1)(\sqrt{z'+1} -1)}~.
\ee
In Fig.~\ref{fig:dip1} the relative amplitude of $C_1$ as a function
of $z$ for different values of $z'$ is shown.  In Fig.~\ref{fig:dip2} we plot
$C_1(z,z)$. It seems to be most promising to measure the dipole at
relatively low redshift. But of course, the redshift must be
sufficiently high such that the peculiar velocities of the supernovae
themselves are not strongly correlated with each other or with our
peculiar motion. Hence the distance of a supernova at $z$ or $z'$
should be sufficient for linear perturbation theory to apply. This is
safely achieved for $z,z'\gsim 0.1$. At $z=z'=0.1$ we have 
$C_1(0.1,0.1) \simeq 105\times C_1^{\mr{CMB}}$, hence an enhancement
of about a factor 100 with respect to the CMB dipole. This factor is
even somewhat 
larger, in a $\Lambda$--dominated cosmology. Through its dependence on
$\apt(z)$,  measuring the
amplitude of this dipole alone can already lead to new observational
constraints on the expansion history of the universe.

\subsection{The higher multipoles}
 
We now want to express the higher $C_\ell$'s in terms of the power spectrum
for the Bardeen potential. We define the Fourier transform
\begin{equation}
\Psi(\eta,\bk)=\int d^3x e^{-i\bk\mathbf{x}}\Psi(\eta,\mathbf{x}) ~.
\end{equation}
We split the deterministic time evolution into a 'transfer function'
$T_k(\eta)$, such that $\Psi(\eta,\bk) =T_k(\eta)\Psi(\bk)$. We
normalize the transfer function such that $\lim_{k\ra 0}T_k(\eta_0) =1$.
The power spectrum $P_\Psi$ of $\Psi(\bk)$ is defined by
\be\label{e:defP}
k^3\langle \Psi(\bk)\Psi^*(\bk')\rangle = (2\pi)^3\de^3(\bk-\bk')P_\Psi(k)
~.
\ee
The $\de^3$-- function is a consequence of statistical homogeneity.
We need to determine the correlation function of $\Psi$ for the 
positions $\bx = \bx_O -\bn(\eta_O-\eta)$ and  $\bx' = \bx_O
-\bn'(\eta_O-\eta')$.
In terms of the power spectrum the correlation function of $\Psi$ and
of its derivatives as they enter in Eq.~(\ref{eq:dL(zS)}) can be written
as (for details see Appendices~\ref{appe1} and \ref{appe2}).

\onecolumngrid

\bea
\langle\Psi(\eta,\bx)\Psi(\eta',\bx')\rangle &=&
\sum_{\ell}\frac{2\ell +1}{4\pi}C_\ell^{(\Psi)}(z,z')P_\ell(\bn\cd\bn')
  \qquad \mbox{with} \nonumber \\
 C_\ell^{(\Psi)}(z,z') &=&
\frac{2}{\pi}\int\frac{dk}{k}T_k(\eta)T_k(\eta')P_\Psi(k)j_\ell(k(\eta_O-\eta))
j_\ell(k(\eta_O-\eta'))  \label{eq:CP} \\
\langle\bn\cd\bnabla\Psi(\eta,\bx)\Psi(\eta',\bx')\rangle &=&
\sum_{\ell}\frac{2\ell +1}{4\pi}C_\ell^{(nd\Psi)}(z,z')P_\ell(\bn\cd\bn')
  \qquad \mbox{with} \nonumber \\
C_\ell^{(nd\Psi)}(z,z') &=& -\frac{2}{\pi}\int dkT_k(\eta)T_k(\eta')
P_\Psi(k) j'_\ell(k(\eta_O-\eta)) j_\ell(k(\eta_O-\eta'))  \label{eq:CP1} \\
\langle n^in^j\dd_i\dd_j\Psi(\eta,\bx)\Psi(\eta',\bx')\rangle &=&
\sum_{\ell}\frac{2\ell +1}{4\pi}C_\ell^{(nndd\Psi)}(z,z')P_\ell(\bn\cd\bn')
  \qquad \mbox{with} \nonumber \\
C_\ell^{(nndd\Psi)}(z,z') &=& \frac{2}{\pi}\int dkkT_k(\eta)T_k(\eta')
P_\Psi(k) j^{\prime\prime}_\ell(k(\eta_O-\eta)) j_\ell(k(\eta_O-\eta'))
\label{eq:CP2} \\
\langle \bnabla^2\Psi(\eta,\bx)\Psi(\eta',\bx')\rangle &=&
\sum_{\ell}\frac{2\ell +1}{4\pi}C_\ell^{(dd\Psi)}(z,z')P_\ell(\bn\cd\bn')
  \qquad \mbox{with} \nonumber \\
C_\ell^{(dd\Psi)}(z,z') &=& \frac{-2}{\pi}\int dkkT_k(\eta)T_k(\eta')
P_\Psi(k) j_\ell(k(\eta_O-\eta)) j_\ell(k(\eta_O-\eta'))
\label{eq:CP3} \\
\langle\bn\cd\bnabla\Psi(\eta,\bx)\bn'\cd\bnabla\Psi(\eta',\bx')\rangle &=&
\sum_{\ell}\frac{2\ell +1}{4\pi}C_\ell^{(nd\Psi nd)}(z,z')P_\ell(\bn\cd\bn')
  \qquad \mbox{with} \nonumber \\
C_\ell^{(nd\Psi nd)}(z,z') &=& \frac{2}{\pi}\int dk kT_k(\eta)T_k(\eta')
P_\Psi(k) j'_\ell(k(\eta_O-\eta)) j'_\ell(k(\eta_O-\eta'))  \label{eq:CP4} \\
\langle\bn\cd\bnabla\Psi(\eta,\bx)\bnabla^2\Psi(\eta',\bx')\rangle &=&
\sum_{\ell}\frac{2\ell +1}{4\pi}C_\ell^{(nd\Psi dd)}(z,z')P_\ell(\bn\cd\bn')
  \qquad \mbox{with} \nonumber \\
C_\ell^{(nd\Psi dd)}(z,z') &=& \frac{2}{\pi}\int dk k^2T_k(\eta)T_k(\eta')
P_\Psi(k) j'_\ell(k(\eta_O-\eta)) j_\ell(k(\eta_O-\eta'))  \label{eq:CP5} \\
\langle\bn\cd\bnabla\Psi(\eta,\bx)n^in^j\dd_i\dd_j\Psi(\eta',\bx')\rangle &=&
\sum_{\ell}\frac{2\ell +1}{4\pi}C_\ell^{(nd\Psi ndnd)}(z,z')P_\ell(\bn\cd\bn')
  \qquad \mbox{with} \nonumber \\
C_\ell^{(nd\Psi ndnd)}(z,z') &=& -\frac{2}{\pi}\int dk k^2T_k(\eta)T_k(\eta')
P_\Psi(k) j^{\prime}_\ell(k(\eta_O-\eta))
j^{\prime\prime}_\ell(k(\eta_O-\eta'))  \label{eq:CP6} \\
\langle n^in^j\dd_i\dd_j\Psi(\eta,\bx) n^{\prime i}n^{\prime j}
\dd_i\dd_j\Psi(\eta',\bx')\rangle &=&
\sum_{\ell}\frac{2\ell +1}{4\pi} C_\ell^{(nndd\Psi nndd)}(z,z')
P_\ell(\bn\cd\bn')   \qquad \mbox{with} \nonumber \\
C_\ell^{(nndd\Psi nndd)}(z,z') &=& \frac{2}{\pi}\int dkk^3T_k(\eta)T_k(\eta')
P_\Psi(k) j^{\prime\prime}_\ell(k(\eta_O-\eta))
j^{\prime\prime}_\ell(k(\eta_O-\eta')) \label{eq:CP7} \\
\langle n^in^j\dd_i\dd_j\Psi(\eta,\bx)\bnabla^2\Psi(\eta',\bx')\rangle &=&
\sum_{\ell}\frac{2\ell +1}{4\pi}C_\ell^{(ndnd\Psi dd)}(z,z')P_\ell(\bn\cd\bn')
  \qquad \mbox{with} \nonumber \\
C_\ell^{(ndnd\Psi dd)}(z,z') &=& -\frac{2}{\pi}\int dk
k^3T_k(\eta)T_k(\eta') P_\Psi(k)
j^{\prime\prime}_\ell(k(\eta_O-\eta)) j_\ell(k(\eta_O-\eta'))  \nonumber \\
\langle \bnabla^2\Psi(\eta,\bx)\bnabla^2\Psi(\eta',\bx')\rangle &=&
\sum_{\ell}\frac{2\ell +1}{4\pi}C_\ell^{(dd\Psi dd)}(z,z')P_\ell(\bn\cd\bn')
  \qquad \mbox{with} \nonumber \\
C_\ell^{(dd\Psi dd)}(z,z') &=& \frac{2}{\pi}\int dk
k^3T_k(\eta)T_k(\eta') P_\Psi(k)
j_\ell(k(\eta_O-\eta)) j_\ell(k(\eta_O-\eta'))~.
\label{eq:CP8} \eea

Using these definitions we can write the correlation function of
the luminosity distance as

\begin{equation}\label{eq:dLCl}
\frac{\langle d_L(z_S,\bn)d_L(z_{S'},\bn')\rangle}{\bar d_L(z_S) \bar
d_L(z_{S'})} ~ = ~ \sum_\ell\frac{2\ell+1}{4\pi}P_\ell(\mathbf{n}\mathbf{n'})
\left(C_\ell^{(1)}+C_\ell^{(2)}+C_\ell^{(3)}+C_\ell^{(4)}+C_\ell^{(5)}\right)
\end{equation}

\twocolumngrid

where $C_\ell^{(i)}$ collects all the contributions to $C_\ell$
which contain integrals of the form $\int dk k^{i-2}...$. The
detailed expressions for the $C_\ell^{(i)}$'s are given in
Appendix~\ref{appe2}. Here we just note that the  term
$C^{(5)}_\ell$ represents the lensing contribution. As we shall see, it
dominates for sufficiently high redshift and sufficiently large
$\ell$. Another important contribution is $C_\ell^{(3)}$ which
contains the peculiar velocity of the emitter, the Doppler term. (It
also includes other contributions which are, however, always
subdominant.) 

The results of this section allow the determination of the
luminosity distance for a given initial spectrum $P_\Psi(k)$ and
given transfer function $T_k(\eta)$. The transfer function, the
conformal time $\eta(z)$, as well as the conformal Hubble
parameter $\apt(z)$ depend crucially on the
cosmological parameters. In a forthcoming paper~\cite{next} we
will present a code to determine the luminosity distance power
spectrum numerically and discuss its dependence on cosmological
parameters. In this work, were we mainly want to present the
method, we approximatively calculate the power spectrum for a
simple case to gain some intuition about the order of magnitude of the
different terms.

\section{Results for a pure CDM universe}\label{sec:CDM}
In this section we approximate the luminosity distance power
spectrum semi-analytically for the simple case of a cold dark matter
(CDM) universe without cosmological constant, $\Om_m =1,~ \Om_\Lambda
=0$. We assume a scale invariant spectrum of initial fluctuations, 
\be
P_\Psi(k) = A(k\eta_0)^{n-1} = A, \qquad n=1~.
\ee
The amplitude $A$ is known
from the Wilkinson Microwave Anisotropy Probe (WMAP) experiment, $A \simeq 10^{-10}$ \cite{WMAP}.

In the radiation dominated past of the universe, the Bardeen potential is
constant on super horizon scales, $k\eta<1$ and oscillates and
decays like $1/a^2 \propto 1/\eta^2$ on sub-horizon scales. During
matter domination, the Bardeen potential is constant~\cite{D94}.
To take this gross behavior into account, we approximate the transfer
function during the matter era by
\begin{equation}
T_k(\eta)T_k(\eta') =T_k^2 \simeq \frac{1}{1 + \beta (k\eta_\mr{eq})^4}~,
\end{equation}
where $\eta_\mr{eq}$ denotes the value of conformal time at
matter and radiation equality. Comparing this rather crude
approximation with the numerical one, which can be found e.g. in
Dodelson's book~\cite{Dod}, we find $\beta \simeq 3\times 10^{-4}$. In
addition, there is a log-correction which comes from the logarithmic
growth of matter perturbations during the radiation era. We shall take
it into account only 
for the dominant term $C_\ell^{(5)}$. Furthermore, we use that during
the matter dominated era $ 4\pi Ga^2(\rho+p) = \frac{3}{2}(\dot a/a)^2
= \frac{3}{2}(2/\eta)^2 =6/\eta^2$.

To determine the power spectrum, we have to perform
integrals over time of the form
\bea
I(f) &=& \int_{\eta_S}^{\eta}d\eta' f(\eta')j_\ell(k(\eta_0-\eta'))
\nonumber \\
&=& \frac{1}{k}\int_x^{x_S}dx' f(\eta_0-x'/k)j_\ell(x')~,
\eea
where we have introduced $x= k(\eta_0-\eta)$. The spherical Bessel
function of order $\ell$ 
is peaked at $x\simeq \ell$. For values much smaller than $\ell$ it is
suppressed like $(x/\ell)^\ell$ and for values much larger that
$\ell$ it oscillates and decays like $1/x$. In our crude approximation,
we neglect contributions to this integral from outside the first peak
and approximate the integral over the first peak by the value of
$f$ at $x=\ell$ multiplied by the area under the peak. This gives
\be \label{eq:I(f)}
I(f) \simeq \frac{1}{k}I_\ell f\left(\eta_0-\frac{\ell}{k}\right)
   \theta\left(k-\frac{\ell}{\eta_0-\eta_S}\right)
\theta\left(\frac{\ell}{\eta_0-\eta}-k\right) ~,
\ee
where $I_\ell$ is the area under the first peak of the Bessel function $j_\ell$
and $\theta$ denotes the Heaviside function, $\theta(x) =0$, if $x\le 0$ and
$\theta(x) =1$, if $x > 0$. Numerically we have found $I_\ell^2 \simeq
1.58/\ell$. 
Most of the resulting integrals over $k$ can either be obtained analytically
in terms of hyper-geometric functions~\cite{AS} or they can be approximated
by the same method. Finally, one $k$-integral contributing to the
Doppler term $C^{(3)}_\ell$
has  to be performed numerically. More details are given in
Appendix~\ref{appe3}. 

We have tested our approximations by comparing them with the numerical
result and have found that we nearly always overestimate the numerical
result, but never by more than a factor of $2$. The approximations are
quite bad at low $\ell\le 5$, but become reasonable later. A fully numerical
evaluation as we shall perform it in~\cite{next}, will probably give a
somewhat smaller result but not by more that a factor of $2$ to $4$. Here, we
are not so much 
interested in numerical accuracy as in qualitative features of the
different contributions to the power spectrum.

\begin{figure}
\centerline{\epsfig{figure=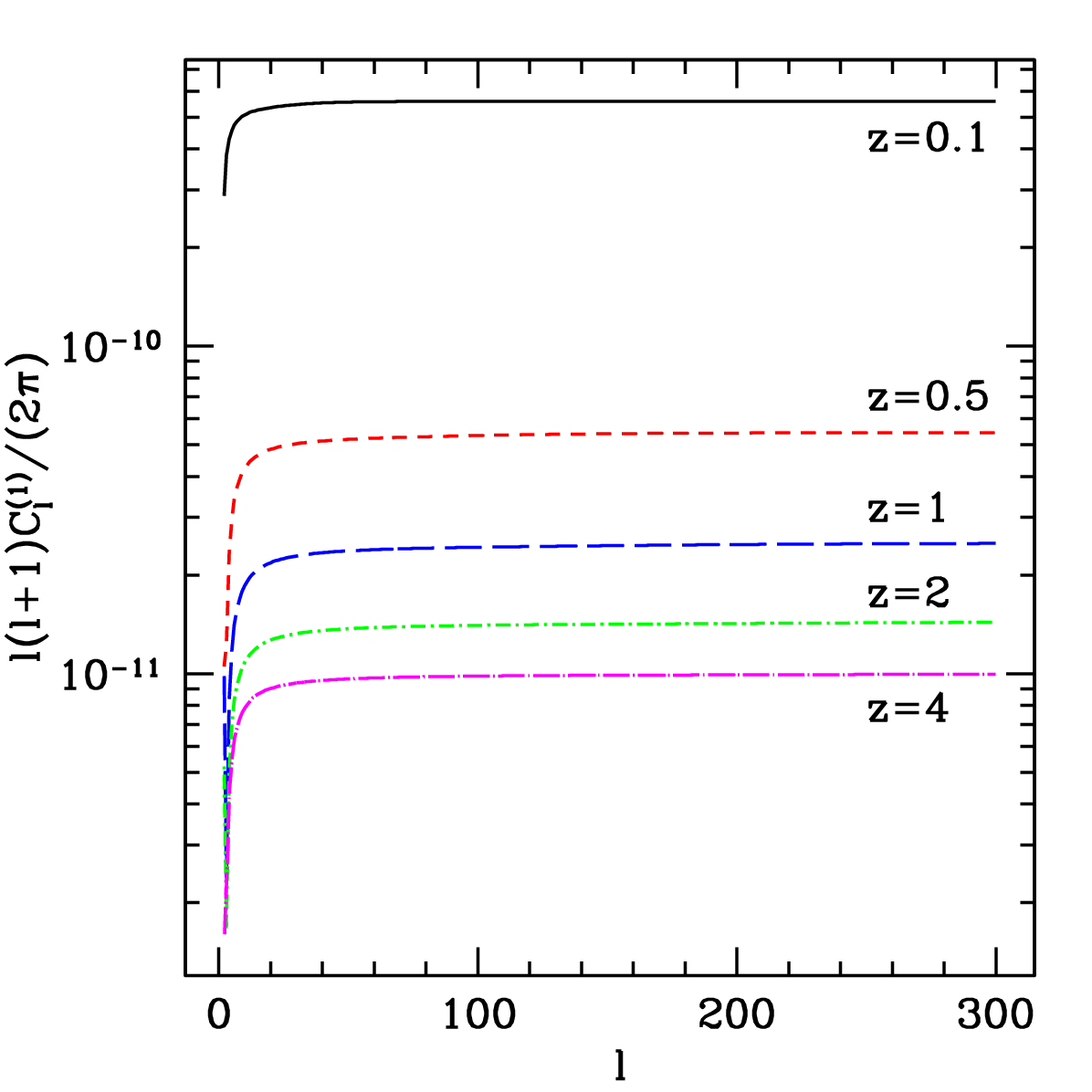,width=6.5cm}}
\caption{ \label{fig:C1} The contribution of the redshift term
  $\ell(\ell+1)C_\ell^{(1)}(z,z)/(2\pi)$ 
  for $z=0.1, 0.5, 1, 2$ and $4$ (from top to bottom).}
\end{figure}

\begin{figure}
\centerline{\epsfig{figure=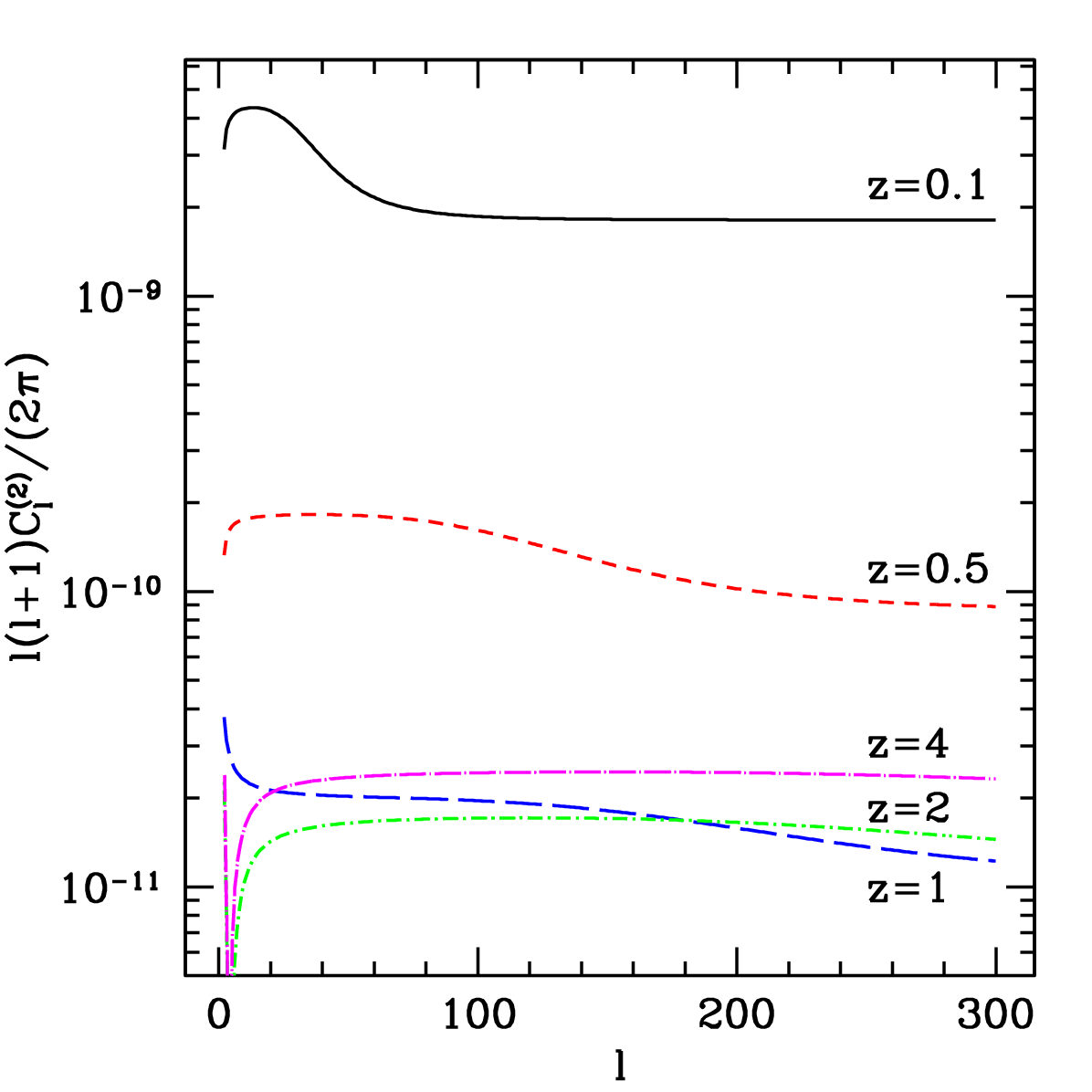,width=6.5cm}}
\caption{ \label{fig:C2} The contribution $C_\ell^{(2)}$. We choose
  the same colors (linestyles) like in Fig.~\ref{fig:C1}. The contributions 
  for $z=0.1, 0.5$ and $1$ are negative while those for $z=2, 4$ pass
  through 0 at low $\ell$, visible as a spike. This may well be due to
  our approximative treatment.}
\end{figure}

\begin{figure}
\centerline{\epsfig{figure=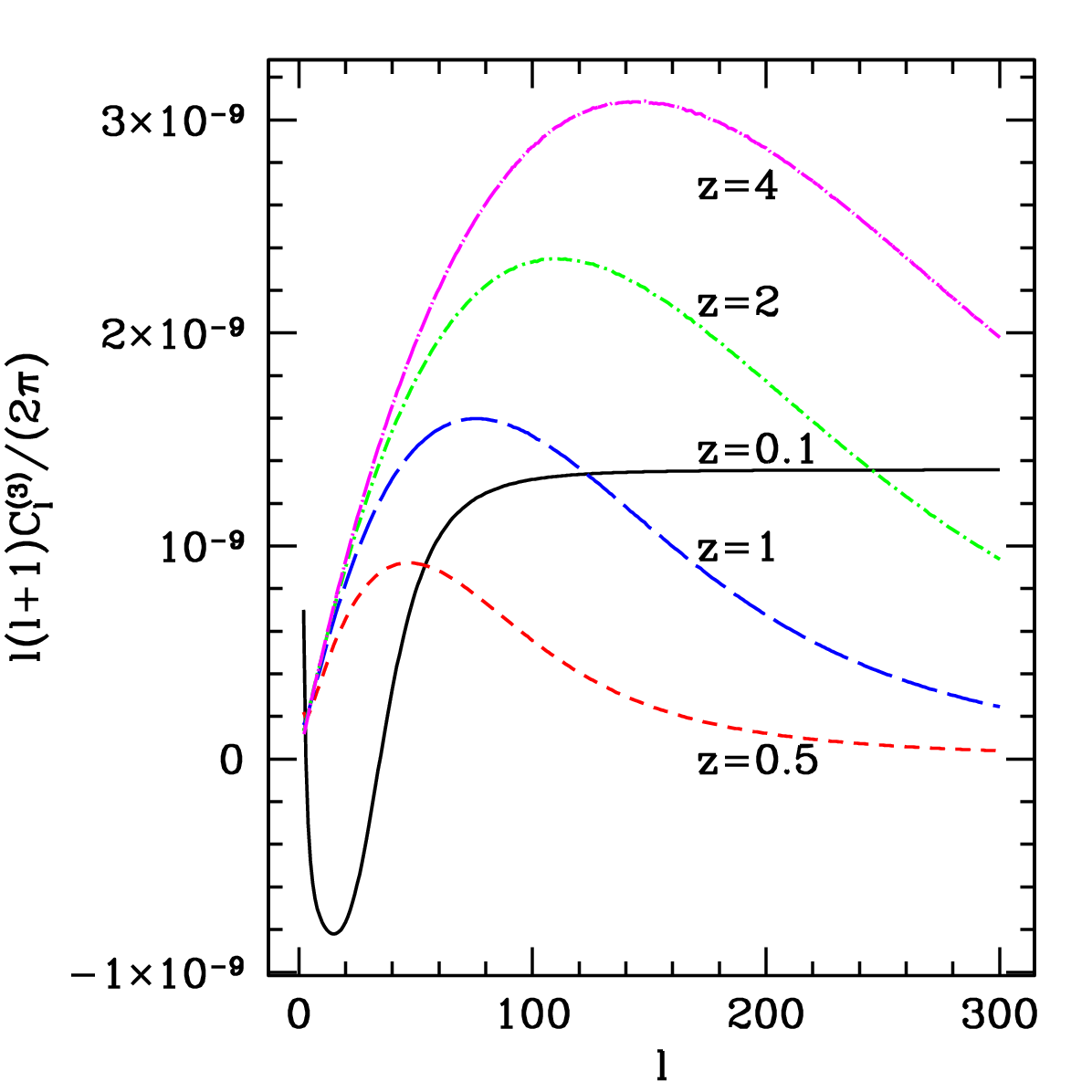,width=6.5cm}}
\caption{ \label{fig:C3} The contribution $\ell(\ell+1)C_\ell^{(3)}(z,z)/(2\pi)$, without the numerical part,
  for $z=4,2,1,0.5$ and $0.1$ (from top to bottom). Note that here we
  have chosen linear as opposed to a log representation.}
\end{figure}

\begin{figure}
\centerline{\epsfig{figure=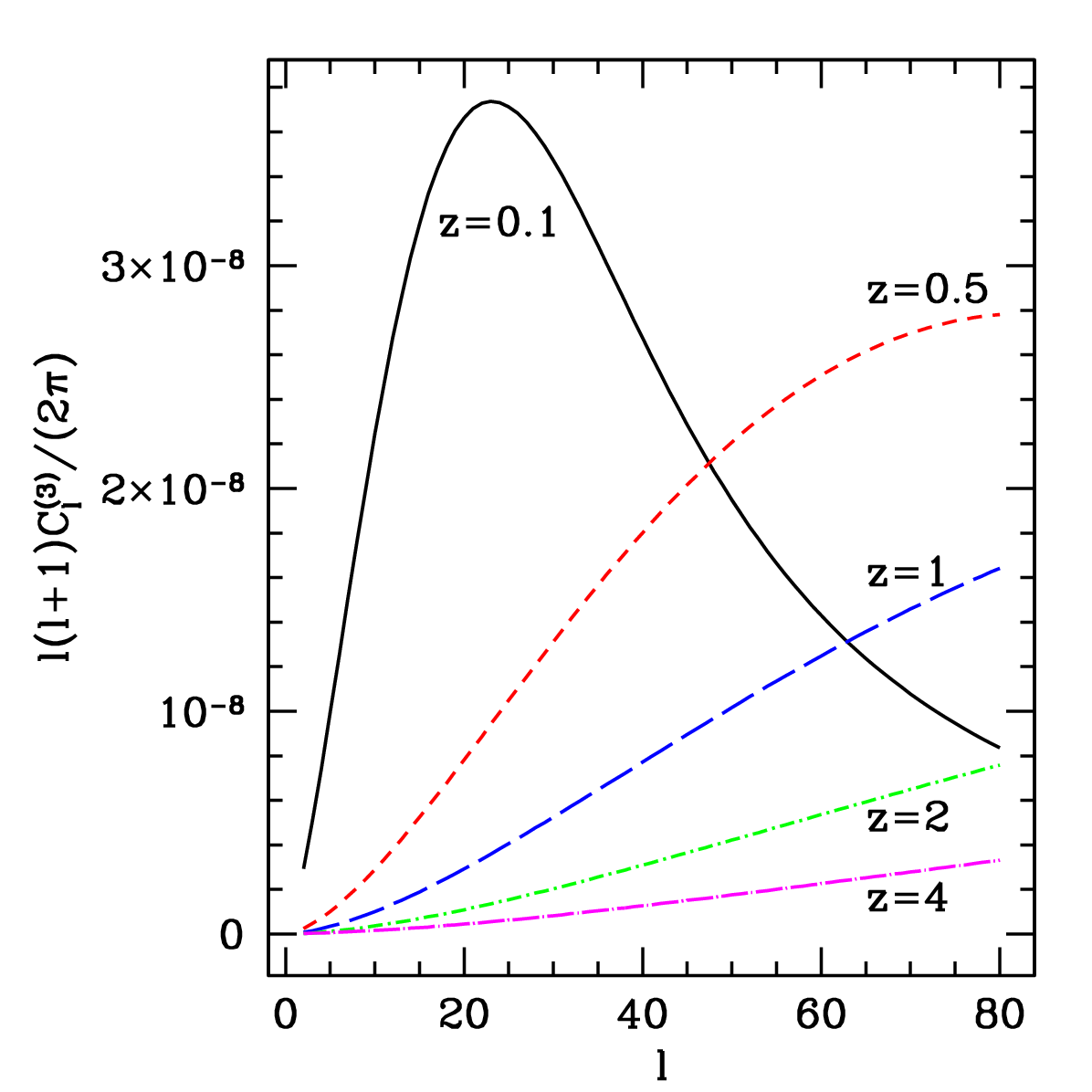,width=6.5cm}}
\caption{ \label{fig:C3num} The Doppler contribution  of
  $\ell(\ell+1)C_\ell^{(3)}(z,z)/(2\pi)$ which has been determined numerically 
  for $z=0.1, 0.5, 1, 2$ and $4 $ (from top to bottom). Our numerical
  code is stable only for $\ell \lsim 80$ and we therefore plot only
  this part of the curve.}
\end{figure}

\begin{figure}
\centerline{\epsfig{figure=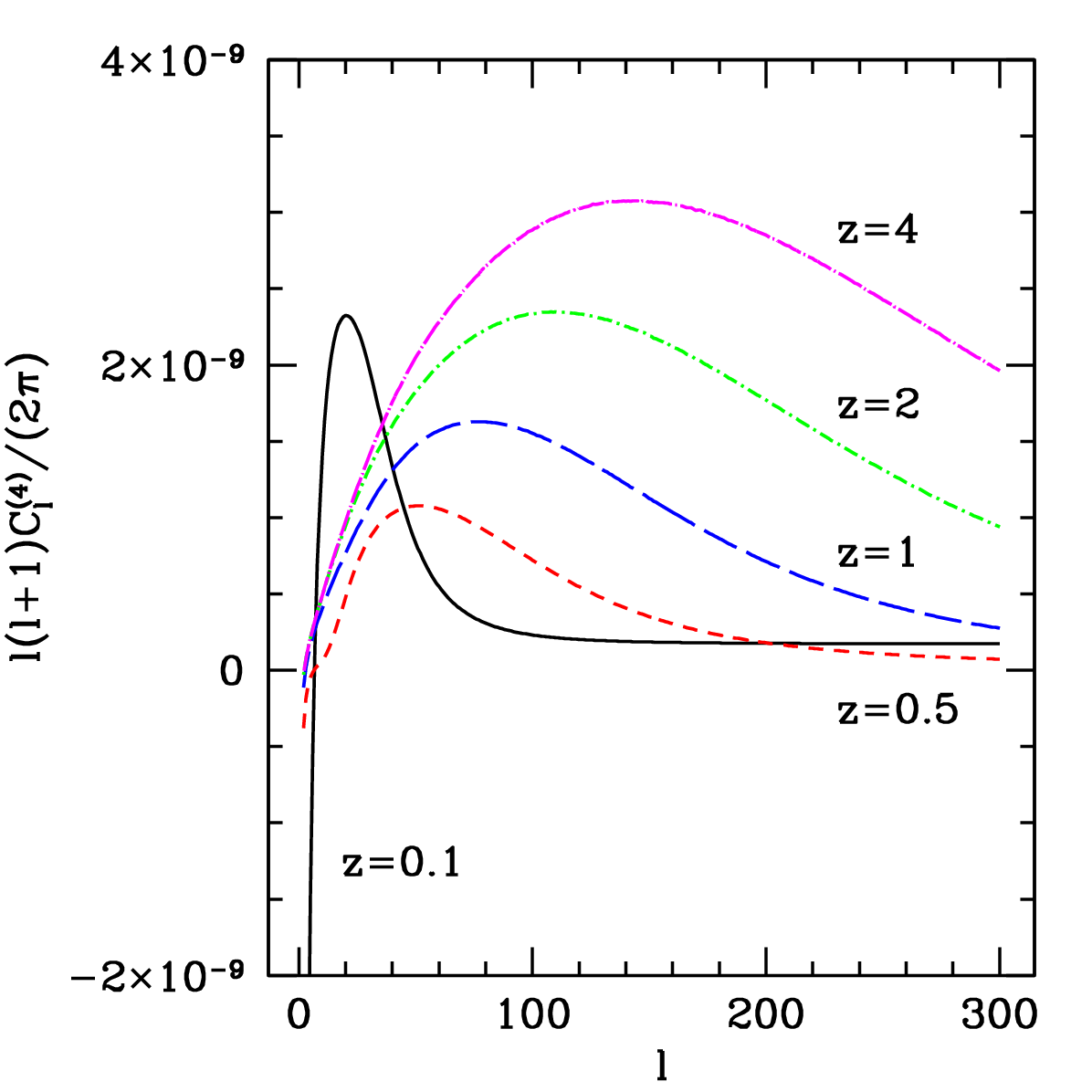,width=6.5cm}}
\caption{ \label{fig:C4} The contribution $\ell(\ell+1)C_\ell^{(4)}(z,z)/(2\pi)$
  for $z= 4,2,1,0.5$ and $0.1$ (from top to bottom).}
\end{figure}

\begin{figure}
\centerline{\epsfig{figure=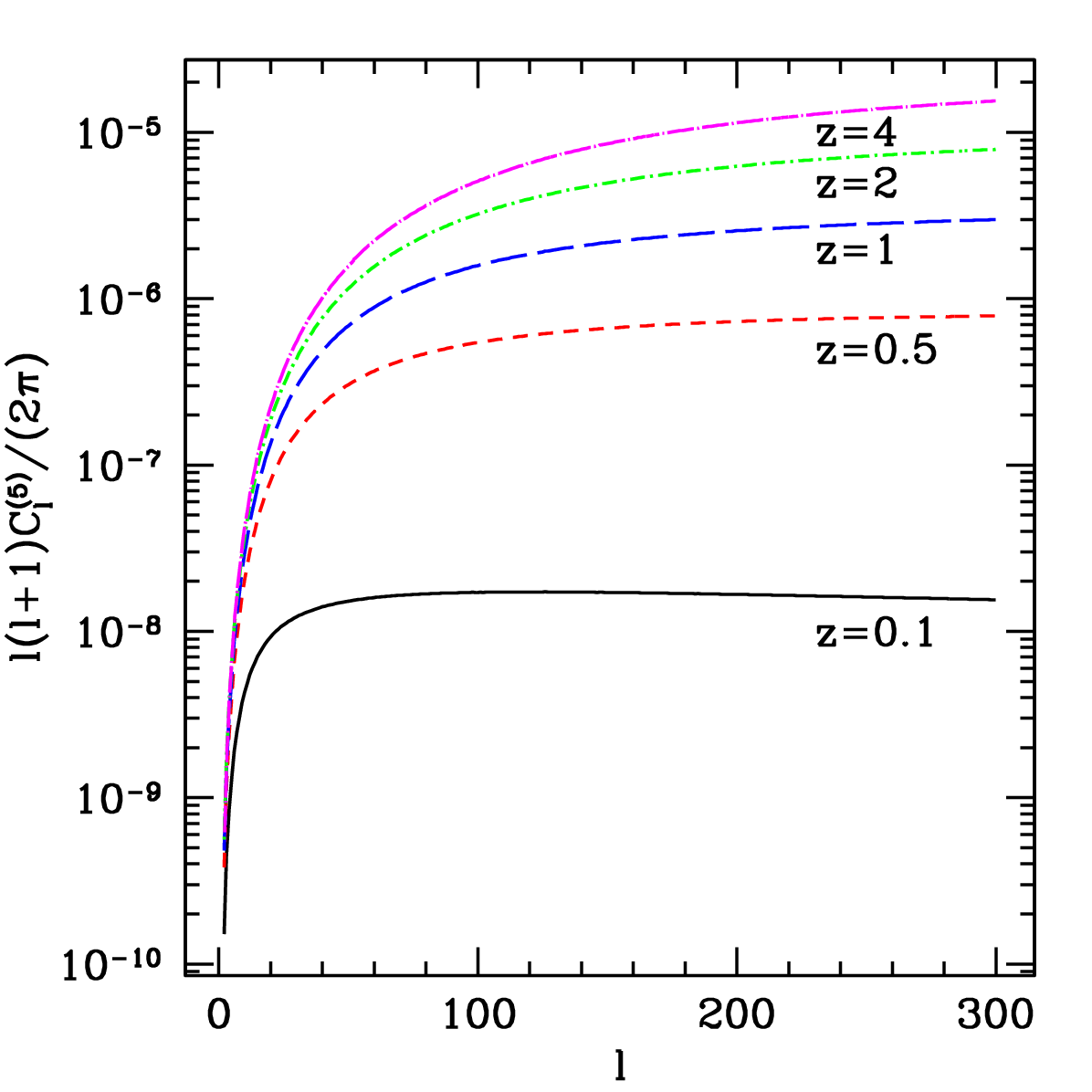,width=6.5cm}}
\caption{ \label{fig:C5} The lensing contribution
  $\ell(\ell+1)C_\ell^{(5)}(z,z)/(2\pi)$ 
  for $z=4,2,1,0.5$ and $0.1$ (from top to bottom). For clarity, we
  have again chosen a log representation in this graph.}
\end{figure}

\begin{figure}
\centerline{\epsfig{figure=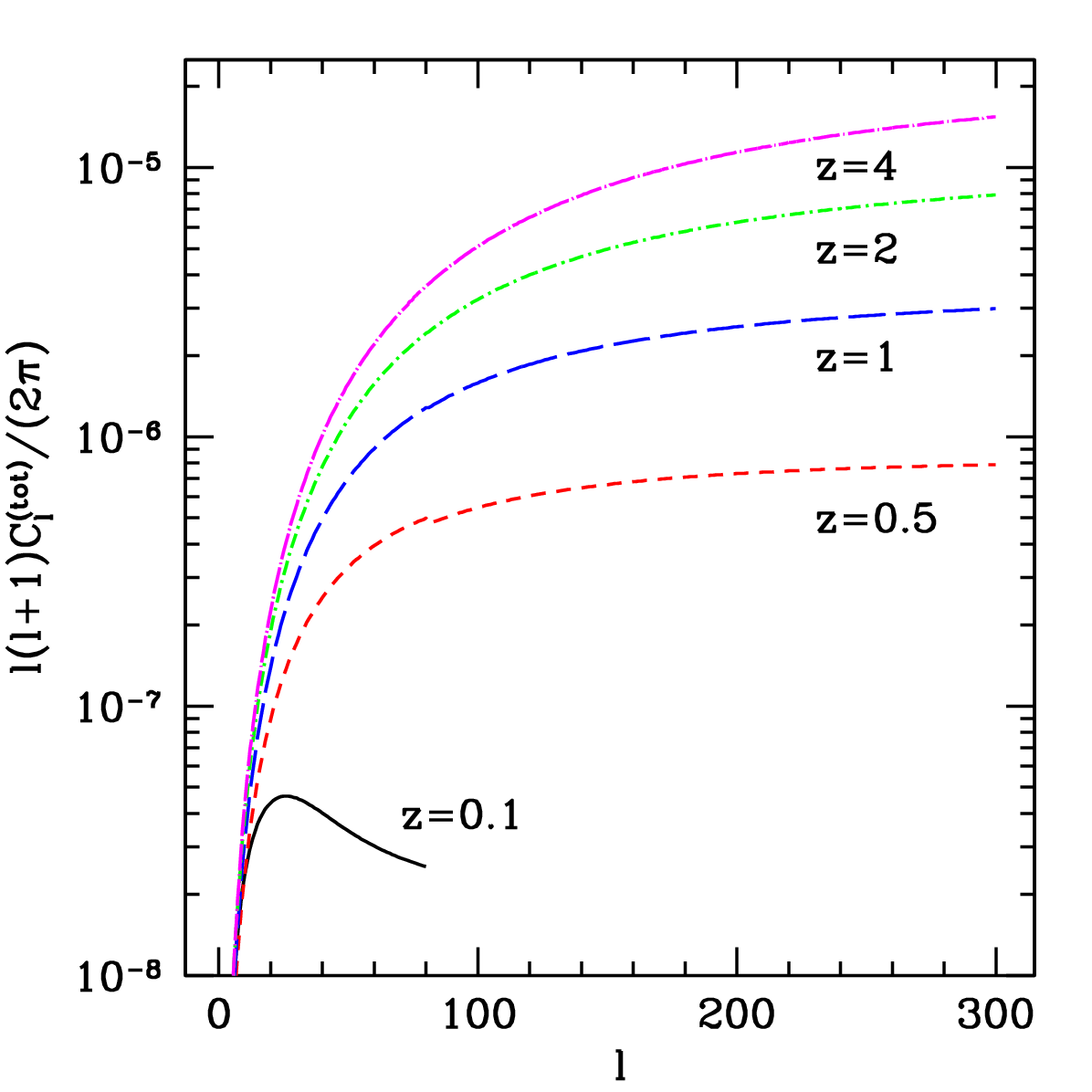,width=6.5cm}}
\caption{ \label{fig:Ctot} The total $\ell(\ell+1)C_\ell(z,z)/(2\pi)$
is shown for $z=4,2,1,0.5$ and $0.1$ (from top to bottom). Note that
for $z>0.1$ it reproduces simply $C_\ell^{(5)}$. For $z=0.1$ the
contribution  of the Doppler part of
$\ell(\ell+1)C_\ell^{(3)}(z,z)/(2\pi)$ is important, which we have
computed only for $\ell\lsim 80$. For clarity, we
  have again chosen a log representation in this graph.}
\end{figure}

\begin{figure}
\centerline{\epsfig{figure=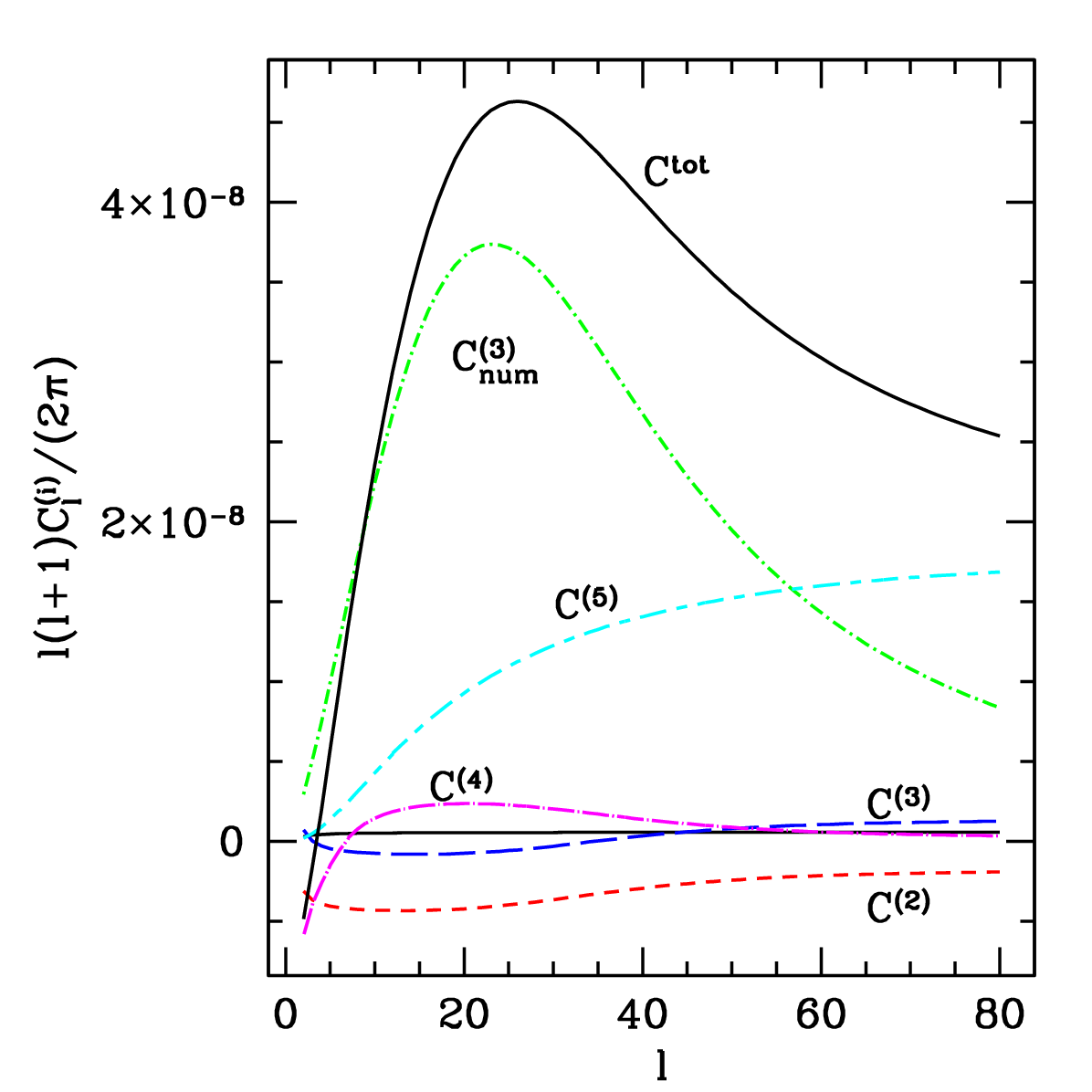,width=7cm}}
\caption{ \label{fig:Ctot80} The different contributions to
  $\ell(\ell+1)C_\ell(z,z)/(2\pi)$  for $z=0.1$ are shown. For this
  low redshift they are all of the same order of magnitude. For low
  $\ell$'s our approximations are not trustable, they even lead to
  negative values for $C_\ell^\mr{tot}$ for $\ell\le 3$.} 
\end{figure}

In Figs.~\ref{fig:C1} to~\ref{fig:C5} we show $\ell(\ell+1)C_\ell^{(i)}(z,z)$
for different values of $z$. For $\ell\gsim 10$, the lensing contribution
$C_\ell^{(5)}$ 
always dominates if $z>0.2$. It is interesting to note that the
different contributions 
do not scale in the same way with $\ell$. Only $C^{(1)}$ and $C^{(2)}$
are scale-invariant with 
\bea
\ell(\ell+1)C^{(1)}_\ell &\simeq &10^{-10}  \\
\ell(\ell+1)C^{(2)}_\ell &\simeq &-10^{-10} ~.
\eea
The other contributions grow up to a redshift dependent maximum
(minimum) from where they decay. They may  become scale invariant at
higher $\ell$, but until $\ell=300$  the
scale invariant piece is only clearly visible for
$z=0.1$. Higher values of $z$ have their maximum contribution at
higher $\ell$ and have not decayed into a scale invariant behavior
 until $\ell=300$.  The lensing contribution $C^{(5)}$ even just grows. For
 $z=0.1$ it does seem to reach a scale invariant plateau, for $z=0.5$
 it seems just to reach the turns over around $\ell=300$. For values
 $z>0.5$ shown in  Fig.~\ref{fig:C5}, the spectrum is simply growing and
has not yet  reached the turn over until $\ell=300$.

The most surprising result is the high amplitude of the lensing
term $C^{(5)}$. Let us  discuss this term in more detail. After performing the
time integrals as outlined above, an integral
$\int dk k^{-1}T_k^2$ from $\ell/(\eta_0-\eta_S)$ to infinity is left. 
If we neglect the log in the transfer function, this amounts to 
\[
\int_{\frac{\ell}{\eta_0-\eta_S}}^\infty \! \frac{dk}{ k}T_k^2 \simeq
\left\{\begin{array}{ll}
\log\left(\frac{\eta_0-\eta_S}{\ell\beta^{1/4}\eta_\mr{eq}}\right) &
\mbox{if} \quad \frac{\ell}{\eta_0-\eta_S} <
\frac{1}{\beta^{1/4}\eta_\mr{eq}} \\
\frac{(\eta_0-\eta_S)^4}{4\ell^4\beta\eta_\mr{eq}^4} & \mbox{else.}
\end{array} \right.
\]
Together with the factor $I_\ell^2\times\ell^2$ from the time
integrations, we obtain a
 $\ell^{-1}$ behavior of $\ell(\ell+1)C_\ell^{(5)}$ at large $\ell$,
which is not seen in Fig.~\ref{fig:C5}. However, when taking into
account  also the log correction, the correct amplitude and scaling with
$\ell$ can be estimated in this way (for more details see
Appendix~\ref{appe2}). 
 
This dominant term comes actually 
from the second derivatives of $\Psi$, hence from the Riemann tensor
which describes the tidal force field, i.e. geodesic deviations. 

If the $k$-integral would not be decaying, $\ell(\ell+1)C_\ell^{(5)}$
would be growing like $\sim \ell^3$. But the integrand becomes small
for fluctuations with wave number smaller than about
 $k_\mr{eq}\equiv 1/(\beta^{1/4}\eta_\mr{eq})$. Therefore
$\ell(\ell+1)C_\ell^{(5)}$  has a (broad) maximum $\ell_{\max} \simeq
k_\mr{eq}(\eta_0-\eta_S)$. Hence $\ell_{\max}$ is increasing with the
source redshift. For $z_S\simeq 1$, hence $ \eta_0-\eta_S
\simeq 0.3\eta_0 \simeq 30\eta_\mr{eq}$ we find  $\ell_{\max} \simeq 250$.
The general expression for a matter dominated universe is
\be
\ell_{\max}(z_S) \simeq 760\times \frac{\sqrt{z_S+1}-1}{\sqrt{z_S+1}} ~.
\ee
Our first important finding is that the tidal force field, represented
by $C_\ell^{(5)}$ totally dominates the final result for redshifts
$z_S \gsim z_{S'} \gsim 0.2$. In a numerical
treatment, where we want to reach a 1\% level accuracy, it is
sufficient to consider only  $C_\ell^{(5)}$ for redshifts $z_S \gsim
z_{S'} \gsim 0.5$. Secondly, na\"\i vely one
would expect a result of the order of $\langle \Psi^2\rangle \simeq A
\simeq 10^{-10}$, but we found nearly $10^{-5}$ 
for supernovae with redshift $z_S\sim 2$. This comes from the fact that
in the time integral for $C^{(5)}$, the fluctuation is multiplied by
the conformal distance  $\eta-\eta_S$. A small angular deviation at
$\eta$ builds up to a large deviation at $\eta_S$ if the distance is
large. Furthermore, we deal  with an integrated effect where even if the
deviation from each fluctuation is similar, more small fluctuations
pile up on the way from the supernovae into the telescope. Even if these are
uncorrelated, we still gain a factor $\sqrt{N}$ by piling them
up. These arguments are somewhat simplistic, but they explain, why the
term with most time integrals and with the factor $(\eta-\eta_S)$ dominates. 

In Fig.~\ref{fig:Ctot} we show the sum   
$$\ell(\ell+1)\left[ \sum_{i} C^{(i)}_\ell(z,z)\right]\frac{1}{2\pi} ~. $$
 For $z>0.1$, the total results are
  indistinguishable from $ C^{(5)}_\ell$ alone. Only for $z=0.1$ all
  terms contribute, especially the numerical part of $ C^{(3)}_\ell$
  dominates. We plot this line only until $\ell=80$ since we have no
  reliable results on the numerical contribution to $C^{(3)}_\ell$ for
  higher values of $\ell$. The different contributions to $C_\ell$ for
  $z=z'=0.1$ are shown in more detail in Fig.~\ref{fig:Ctot80}.
 
\begin{figure}[ht]
\centerline{\epsfig{figure=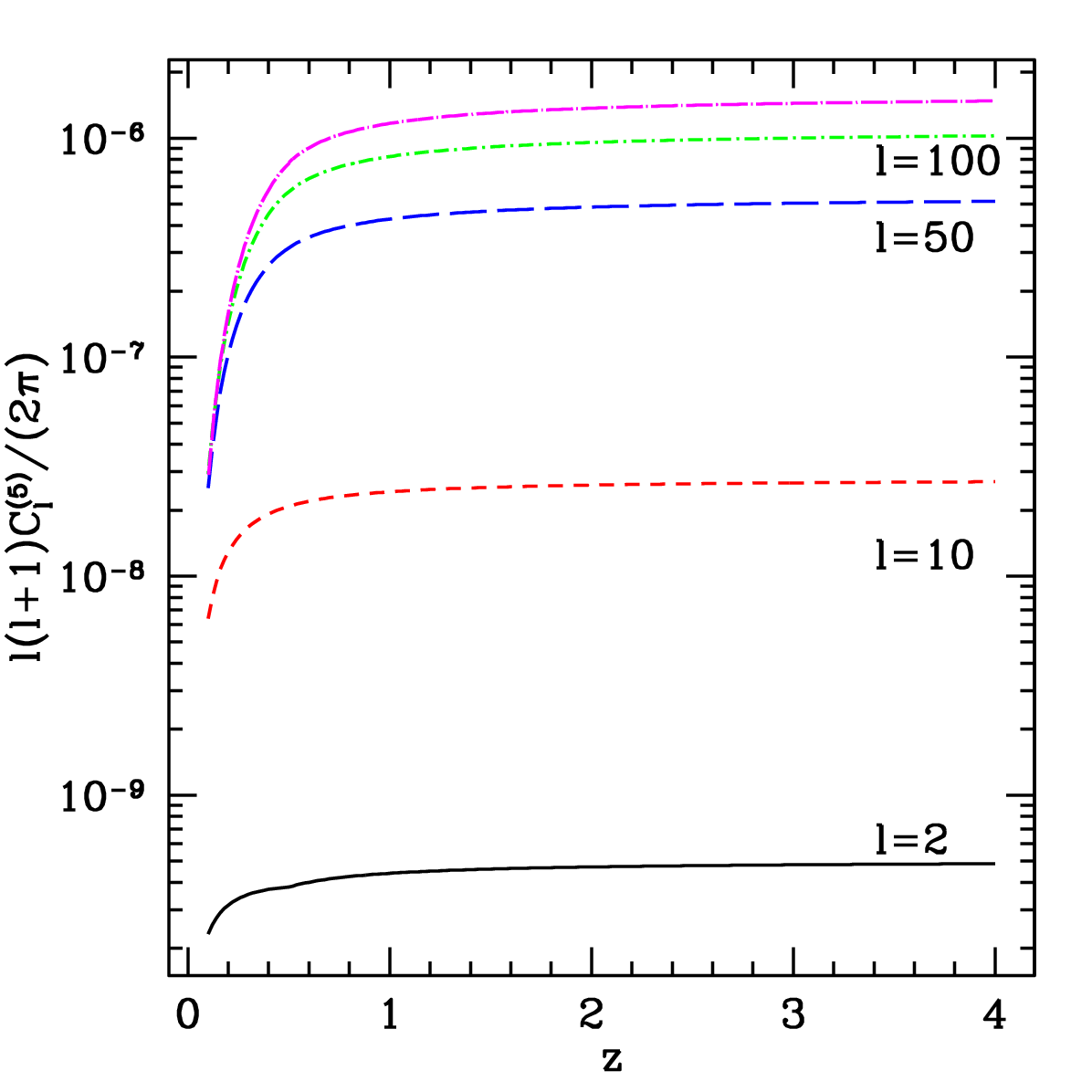,width=7cm}}
\caption{ \label{fig:C5el} The contribution
  $\ell(\ell+1)C_\ell^{(5)}(z,z')/(2\pi)$ is shown as a function of
  $z$ with $z'$ fixed to $z'=0.5 $ and $\ell =200,100,50,10$ and $2$
  (from top to 
  bottom). Above $\ell \simeq 50$, the $\ell$--dependence of the
  result becomes weak as expected. For $z>0.1$ this represents
  actually also the total contribution to $C_\ell$.}
\end{figure}

\begin{figure}[ht]
\centerline{\epsfig{figure=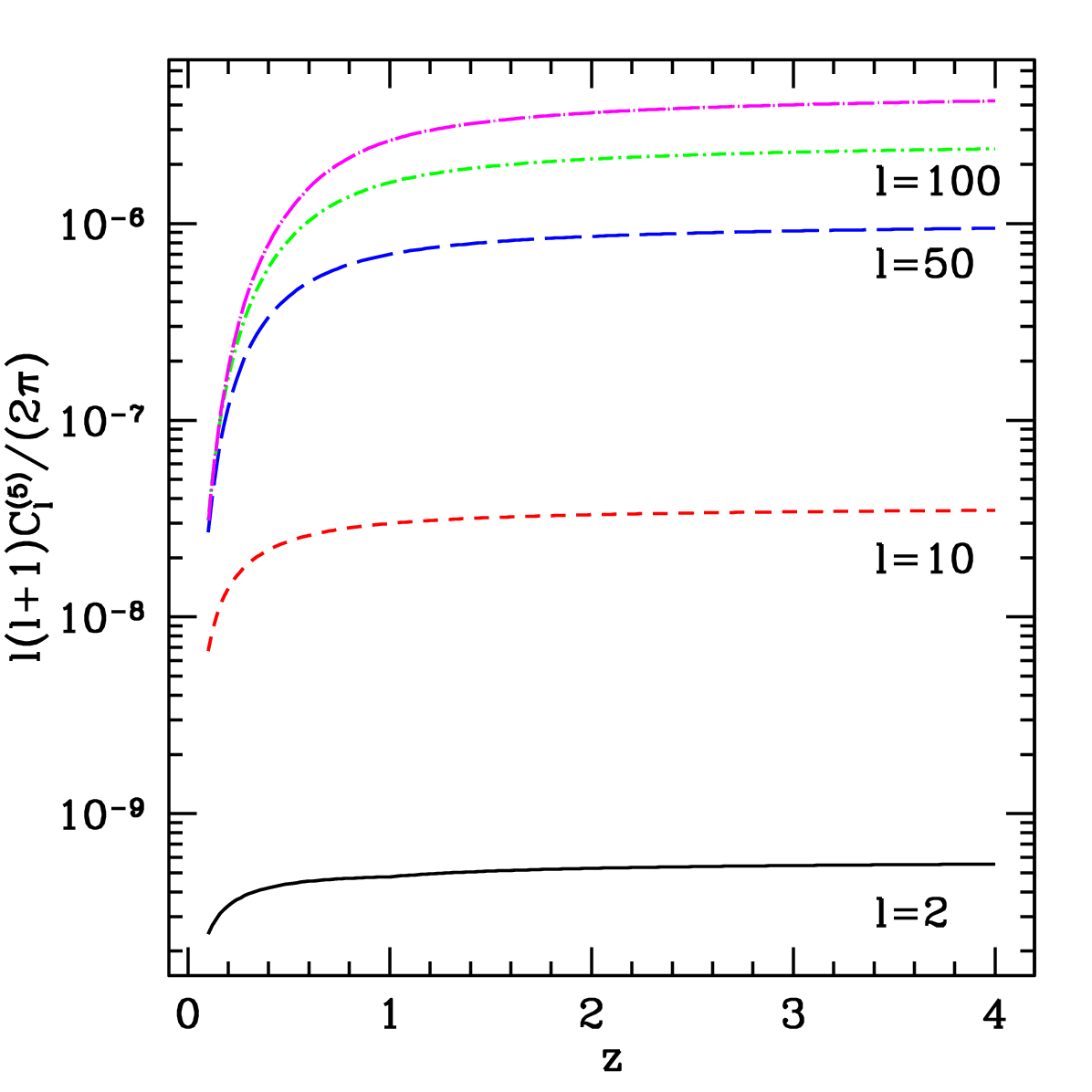,width=7cm}}
\caption{ \label{fig:C5elbis} Like Fig.~\ref{fig:C5el}, but for $z'=1 $.}
\end{figure}

\begin{figure}[ht]
\centerline{\epsfig{figure=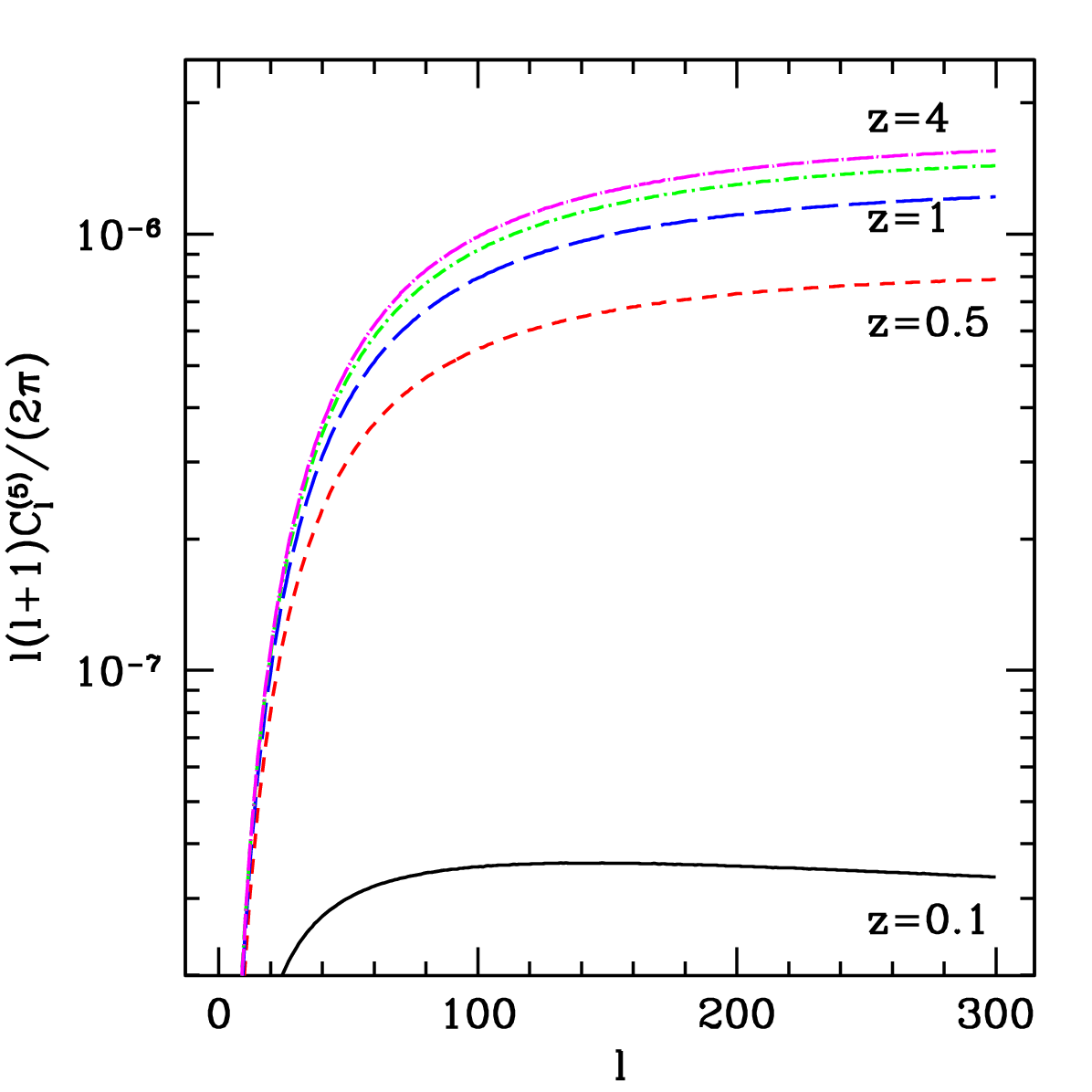,width=7cm}}
\caption{ \label{fig:C5z1z2} The contribution
  $\ell(\ell+1)C_\ell^{(5)}(z,z')/(2\pi)$ 
  is shown for $z'= 0.5 $ and $z=4, 2, 1, 0.5$ and $0.1$ (from top to
  bottom). Again, for $z\neq 0.1$ this result is equivalent to the
  full $C_\ell$.}
\end{figure}

It is also interesting to study the behavior of $C_\ell(z,z')$ for
fixed $z'$ as a function of $z$, and for fixed $z\neq z'$ as a
function of $\ell$. We show this behavior in 
Figs.~\ref{fig:C5el} to \ref{fig:C5z1z2}. Somewhat surprisingly
$C_\ell(z,z')$ shows no peak at $z=z'$. It is therefore not
problematic to include relatively large bins $\De z$ in a study of
$C_\ell(z,z)$. 

\section{Conclusions and outlook}
\label{sec:con}
In this work we have determined the correlation function of the
luminosity distance fluctuations. We have found that at redshifts
$z\ge 0.2$, the result is dominated entirely by the 'lensing term'
$\langle |\De\Psi|^2\rangle $ which is proportional to the 
density fluctuation. Geometrically it comes from the term
$A_i^j =R^j_{\mu\nu i}k^\mu k^\nu$  i.e. the Riemann tensor. Hence
this contribution is due to the tidal force field. We have seen that
it is dominated by fluctuations of the size $\la\simeq \eta_\mr{eq}$
which enter the horizon at matter radiation equality. These
fluctuations have not been damped during the radiation era, but they
are the smallest and therefore the most numerous which have not
suffered damping. Their effect can therefore add up most along the path of the
photon.

We have found that within linear perturbation theory, the $d_L$-power
spectrum is nearly 5 orders of 
magnitude larger than the CMB anisotropy power spectrum! But
nevertheless, the fluctuations obtained within linear theory are
still much smaller than 1. We have also seen that small scale
fluctuations do not significantly contribute to the $C_\ell$'s for low
$\ell$'s. i.e. on large scales. 
This indicates that they cannot change
the observed $d_L(z)$ 
by factors of order unity, which would be needed to mimic accelerated
expansion in a matter dominated universe. 
Also the variance, i.e., the typical deviation of a given luminosity
distance $d_L(\bn, z)$ from the mean, which is dominated by small
scale fluctuations (the lensing contribution) is
$$ \overline d_L(z)^{-2}\langle d_L(\bn,z)^2\rangle =\frac{1}{4\pi} 
  \sum_\ell (2\ell+1)C_\ell \simeq 10^{-5} \ll 1~. $$
Our findings thus indicate that
the explanation of accelerated expansion put forward in~\cite{Kolb} is
probably not realized. Of course we have not taken into account the
change of the transfer function due to nonlinearities. To determine
this effect more precisely we would have to take into account the
non-linearities, especially in the integral for $C_\ell^{(5)}$.

We suggest that the newly derived luminosity distance power spectrum
given by the $C_\ell(z_S,z_{S'})$ can be used as a new observational
tool to determine cosmological parameters. For 1\% accuracy of the
fluctuations at $z_S\gsim 0.5$, only  $C_\ell^{(5)}$ has to be
taken into 
account and therefore the numerical complexity of the problem seems
to be quite moderate. In a future paper~\cite{next} we shall
investigate the possibilities to measure $C_\ell(z_S,z_{S'})$ with
the supernovae searches which are presently under way or in planning.

\acknowledgments
We thank Martin Kunz, Dominik Schwarz and Lam Hui for useful and stimulating
discussions. We thank Marc-Olivier Bettler for his help with a
figure. We are grateful to Enea Di Dio and Norbert Straumann for pointing out an error in the published 
version which is corrected here. We acknowledge financial support from the Swiss National Science foundation.

\appendix
\section{Christoffel symbols and the Riemann tensor of scalar
  perturbations in non-expanding spacetime}\label{ap:geo}
Here we write down the Christoffel symbols and  Riemann tensor for
the metric
\[
 g_{\mu\nu}dx^\mu dx^\nu =  -(1+2\Psi)d\eta^2
+(1-2\Psi)\ga_{ij}dx^idx^j
\]
to first order in the gravitational potential $\Psi$.
\bea
\Ga^0_{00} &=& \dot\Psi \\
\Ga^0_{0i} &=& \dd_i\Psi \\
 \Ga^i_{00} &=& \dd^i\Psi \\
\Ga^j_{0i} &=& -\de_i^j\dot\Psi \\
\Ga^0_{ij} &=& -\ga_{ij}\dot\Psi \\
\Ga^i_{jm} &=& -\de_j^i\dd_m\Psi - \de_m^i\dd_j\Psi + \ga_{jm}\dd^i\Psi 
\eea

\bea
R^0{}_{00j} &=& 0 \\
R^0{}_{0ij} &=& 0 \\
R^0{}_{i0j} &=& -\nabla_i\nabla_j\Psi -\ga_{ij}\ddot \Psi \\
R^0{}_{ijm} &=& \ga_{ij}\nabla_m\Psi -\ga_{im}\nabla_j \dot\Psi \\
R^i{}_{00j} &=& -\nabla^i\nabla_j\Psi -\de^i_j\ddot \Psi \\
R^i{}_{0jm} &=& \de^i_j\nabla_m\dot\Psi -\de^i_m\nabla_j \dot\Psi \\
R^i{}_{j0m} &=& -\de^i_m\nabla_j\dot\Psi +\ga_{jm}\nabla^i \dot\Psi \\
R^i{}_{jmn} &=& -\left[\de^i_n \nabla_j\nabla_m - \de^i_m \nabla_j\nabla_n 
   + \right. \nonumber \\
 && \left. \ga_{jm}\nabla^i\nabla_n -   \ga_{jn}\nabla^i\nabla_m \right]\Psi~. 
\eea
Here $\nabla_i$ denotes the covariant derivative w.r.t the metric
$\ga_{ij}$.

\onecolumngrid

\section{The derivation of Eq.~(\ref{eq:dL(zS)2})}\label{ap:lit}

We first re-introduce the velocity of the source $\bv_S$ and we
collect all terms which contain spatial derivatives of the form $n^i\dd_i\Psi$
at the end. This brings~(\ref{eq:dL(zS)}) into the form (we dismiss the
tilde in this appendix)

\begin{eqnarray}
{d}_L(z_S,\bn)&=&(1+z_S)\Bigg\{(\eta_O-\eta_S)
+\frac{1}{\apt_S}\left(\Psi_O -\bv_O\cd\bn\right)
- \left(\eta_O-\eta_S +  \apt_S^{-1}\right)\Psi_S- \left(\eta_O-\eta_S
-  \apt_S^{-1}\right)\bv_S\cd\bn 
\nonumber\\ && 
 + 2\int_{\eta_S}^{\eta_O}d\eta \Psi
- \int_{\eta_S}^{\eta_O}d\eta
\int_{\eta_S}^{\eta}d\eta' (\eta'-\eta_S)\nabla^2\Psi \nonumber\\ && 
  -\frac{2}{\apt_S}
\int_{\eta_S}^{\eta_O}\! \hspace{-2mm} d\eta \bn\cd\bnabla\Psi
+2\int_{\eta_S}^{\eta_O}\! \hspace{-2mm} d\eta
\int_{\eta_S}^{\eta}\!\!  d\eta'\bn\cd\bnabla \Psi
+ \int_{\eta_S}^{\eta_O}\! \hspace{-2mm}d\eta
\int_{\eta_S}^{\eta}\! \hspace{-2mm} d\eta'
(\eta'-\eta_S)n^in^j\dd_i\dd_{j}\Psi \Bigg\} 
 ~.  \label{eq:dL(zS)Ap}
\end{eqnarray}
Now we use $$\frac{d\Psi}{d\eta} = \dot\Psi +\bn\cd\bnabla\Psi$$ to
convert all derivatives of the form $\bn\cd\bnabla\Psi$ into time
derivatives.
This leads to
\begin{eqnarray}
{d}_L(z_S,\bn)&=&(1+z_S)\Bigg\{(\eta_O-\eta_S)
-\frac{1}{\apt_S}\left(\Psi_O +\bv_O\cd\bn\right)
+ \left(-2(\eta_O-\eta_S) +  \apt_S^{-1}\right)\Psi_S- \left(\eta_O-\eta_S
-  \apt_S^{-1}\right)\bv_S\cd\bn 
\nonumber\\ && 
 +(\eta_O-\eta_S)\Psi_O + 2\int_{\eta_S}^{\eta_O}d\eta \Psi
- \int_{\eta_S}^{\eta_O}d\eta
\int_{\eta_S}^{\eta}d\eta' (\eta'-\eta_S)\nabla^2\Psi
\nonumber\\ && 
  +\frac{2}{\apt_S}
\int_{\eta_S}^{\eta_O}\! \hspace{-2mm}d\eta \dot\Psi
-2\int_{\eta_S}^{\eta_O}\! \hspace{-2mm}d\eta  (\eta-\eta_S)\dot\Psi
+\int_{\eta_S}^{\eta_O}\! \hspace{-2mm} d\eta
\int_{\eta_S}^{\eta}\!\!  d\eta'(\eta'-\eta_S)\ddot\Psi  \Bigg\} ~.
 \label{eq:dL(zS)Ap2}
\end{eqnarray}
Via  integration by parts we can now convert the double
  integrals over time into single integrals. For this we use
  that for a regular function $f(\eta)$ integrating by parts
  $\int_{\eta_S}^{\eta_O}\!\! d\eta(\eta-\eta_S)^2f(\eta)$ gives
$$ \int_{\eta_S}^{\eta_O}\! \hspace{-2mm} d\eta
\int_{\eta_S}^{\eta}\!\!  d\eta'(\eta'-\eta_S)f(\eta') = 
\int_{\eta_S}^{\eta_O}\! \hspace{-2mm} d\eta
(\eta-\eta_S)(\eta_O-\eta)f(\eta)~.$$
Using this in the two double integrals above we obtain Eq.~(\ref{eq:dL(zS)2}).
\vspace{0.3cm}

\twocolumngrid

\section{The power spectrum} \label{appe1}
We use the Fourier transform convention
\bea
\Psi(\bk) &=& \int d^3x e^{-i\bk\cd\bx}\Psi(\bx)~,  \\
\Psi(\bx) &=& \frac{1}{(2\pi)^3}\int d^3k e^{i\bk\cd\bx}\Psi(\bk) ~.
\eea
The time evolution of the Bardeen potential is given by the transfer
function, $\Psi(\bk,\eta) = T_k(\eta)\Psi(\bk)$, which is normalized
such that $\Psi(\bk,\eta_0) \ra \Psi(\bk)$ for $k\ra 0$. Since the
Bardeen potential is constant on very large scales, this identifies
$\Psi(\bk)$ also with the Bardeen potential right after
inflation.  The correlation function 
\[
\zeta_\Psi(|\bx-\by|) \equiv \langle \Psi(\bx)\Psi(\by) \rangle
\]
depends only on the distance $|\bx-\by|$, so that we obtain
\bea
\langle \Psi(\bk,\eta)\Psi^*(\bk',\eta') \rangle = \qquad\qquad && \nonumber\\
T_k(\eta)T_{k'}(\eta') \int d^3xd^3y \zeta_\Psi(|\bx-\by|)
e^{-i\bk\cd\bx+i\bk'\cd\by}   &&
\nonumber \\  \label{eq:PPsi}
= T_k(\eta)T_{k'}(\eta')k^{-3}P_\Psi(k) (2\pi)^3\de^3(\bk-\bk')~, &&
\eea
where we have introduced the power spectrum
\be 
P_\Psi(k) = k^3\int d^3z\zeta_\Psi(\bz)e^{-i\bk\cd\bz} ~. 
\ee
It is easy to verify that this definition is consistent with the one given in
Eq.~(\ref{e:defP}). 
 
Standard inflationary scenarios give $P_\Psi \simeq
A(k\eta_0)^{n-1}$ with $n\simeq 1$. From WMAP and other measurements
of CMB anisotropies we have $A\sim 10^{-10}$. We first want to
determine the correlation of the Bardeen potential at positions
$\bx=\bx_O-\bn(\eta_O-\eta)$ and $\bx'=\bx_O-\bn'(\eta_O-\eta')$. With the
above we have
\bea
\langle \Psi(\eta,\bx)\Psi(\eta',\bx') \rangle =  \qquad\qquad\qquad\qquad
 && \nonumber \\
 \frac{1}{(2\pi)^6}\int d^3kd^3k' T_k(\eta)T_{k'}(\eta')\langle 
 \Psi(\bk)\Psi^*(\bk')\rangle && \nonumber \\
 e^{-i\bk\cd\bn(\eta_O-\eta)}
 e^{+i\bk'\cd\bn'(\eta_O-\eta')}~.\qquad  && 
\eea

Using the identity (see e.g.~\cite{AS})
\be
 e^{i\bk\cd\bn(\eta_O-\eta)} = \sum_{\ell} (2\ell+1) i^{\ell}
 j_\ell(k(\eta_O-\eta)) P_\ell(\hat\bk\cd\bn)
\ee
and Eq.~(\ref{eq:PPsi})
we obtain
\bea
\langle \Psi(\eta,\bx)\Psi(\eta',\bx') \rangle = \qquad\qquad\qquad
&& \nonumber \\ 
\frac{1}{(2\pi)^3}\sum_{\ell\ell'}(2\ell+1)(2\ell'+1)i^{\ell-\ell'} \int
    \frac{dk}{k}T_k(\eta)T_k(\eta')\Bigg[ P_\Psi(k) && \nonumber \\ 
 j_\ell(k(\eta_O\! -\! \eta)) j_{\ell'}(k(\eta_O\! -\! \eta'))
\int d\Om_{\hat\bk}    P_\ell(\hat\bk\cd\bn) 
    P_{\ell'}(\hat\bk\cd\bn')\Bigg]  =   && \nonumber \\
\frac{1}{2\pi^2}\sum_{\ell}(2\ell+1)  P_\ell(\bn\cd\bn')\int
\frac{dk}{k}P_\Psi(k)j_\ell(k(\eta_O-\eta))\times~  && \nonumber \\
 j_{\ell}(k(\eta_O-\eta')) \qquad = \qquad \qquad  && \nonumber \\
\sum_{\ell}\frac{2\ell+1}{4\pi}C^{(\Psi)}_\ell(z,z')P_\ell(\bn\cd\bn')
~, \qquad\qquad
\eea
where we have used Eq.~(\ref{eq:CP}) for the last equal sign.
Here $\hat\bk$ is the unit vector in direction $\bk$ and
$d\Om_{\hat\bk}$ denotes the integral over the sphere of
$\bk$--directions. 

In the same way one derives Eqs.~(\ref{eq:CP1}) to
(\ref{eq:CP8}). Each factor $i\bn\cd\bk$ can be written as a derivative
w.r.t $\eta_O-\eta$ of the exponential and therefore replaces
$j_{\ell}(k(\eta_O-\eta))$ by $-kj_{\ell}'(k(\eta_O-\eta))$.  The
Laplacian simply corresponds to a factor $-k^2$.  

\onecolumngrid

\section{Details for the power spectrum}  \label{appe2}

In this appendix we write down in detail the expressions for the
$C_\ell^{(i)}$'s used in this paper.

As mentioned in Section~\ref{sec:PS} the power spectrum
of the luminosity distance can be split in five different
parts containing $k$-integrals of different powers,

$C_\ell^{(1)}$ contains the integrals of the form $\int \frac{dk}{k}$,
represents the redshift and parts of the integrated contributions.

$C_\ell^{(2)}$ contains the integrals of the form $\int dk$,
represents the correlation of the Doppler term with the terms in $C^{(1)}$.

$C_\ell^{(3)}$ contains the integrals of the form $\int dk\cdot k$,
represents the Doppler term and some (subdominant)
integrated terms.

$C_\ell^{(4)}$ contains the integrals of the form $\int dk\cdot k^2$,
is dominated by the correlation of the Doppler term with the lensing
contribution. 

$C_\ell^{(5)}$ contains the integrals of the form $\int dk\cdot k^3$,
represents the lensing term.

From Eqs.~(\ref{eq:CP}) to~(\ref{eq:CP8}) and the
expression~(\ref{eq:dL(zS)}) 
for the luminosity distance we obtain the following expressions
for the $C_\ell^{(i)}$'s

\begin{eqnarray}
C_\ell^{(1)}&=&\frac{2}{\pi}\int
\frac{dk}{k}P_\Psi(k)\left[\frac{2}{\eta_O-\eta_S}\int_{\eta_S}^{\eta_0}d\eta
  T_k(\eta)j_\ell(k(\eta_0-\eta)) -
  \left(1+\frac{1}{\apt_S(\eta_O-\eta_S)}\right)
  T_k(\eta_S)j_\ell(k(\eta_O-\eta_S))\right]\times \nonumber \\
&& \left[\frac{2}{\eta_O-\eta_{S'}}\int_{\eta_{S'}}^{\eta_0}d\eta
  T_k(\eta)j_\ell(k(\eta_0-\eta)) -
  \left(1+\frac{1}{\apt_{S'}(\eta_O-\eta_{S'})}\right)
  T_k(\eta_{S'})j_\ell(k(\eta_O-\eta_{S'}))\right] ~.
\end{eqnarray}

\begin{eqnarray}
C_\ell^{(2)}&=&-\frac{4}{\pi}\int dk
P_\Psi(k)
 \Bigg[\frac{1}{3\apt_S}\left(1-\frac{1}{\apt_S(\eta_O-\eta_S)}\right)
\left(T_k(\eta_S)+\apt_S^{-1}\dot{T}_k(\eta_S)\right)j'_\ell(k(\eta_0-\eta_S))
\nonumber\\  && \qquad
-\frac{1}{\apt_S(\eta_O-\eta_S)} \int_{\eta_S}^{\eta_0}d\eta
T_k(\eta)j'_\ell(k(\eta_0-\eta)) + 
\frac{1}{\eta_0-\eta_S} \int_{\eta_S}^{\eta_0}d\eta
\int_{\eta_S}^{\eta}d\eta' T_k(\eta')
  j'_\ell(k(\eta_0-\eta'))\Bigg]\times
\nonumber\\
&&   \Bigg[ \frac{2}{\eta_0-\eta_{S'}}\int_{\eta_{S'}}^{\eta_0}d\eta
T_k(\eta)j_\ell(k(\eta_0-\eta)) -
\left(1+\frac{1}{\apt_{S'}(\eta_O-\eta_{S'})} \right)T_k(\eta_{S'})
j_\ell(k(\eta_0-\eta_{S'})) \Bigg]
 \nonumber\\  && \qquad\qquad
\hspace{0.5 cm} +\hspace{0.5 cm}
\eta_S \Leftrightarrow \eta_{S'} ~.
\end{eqnarray}

\begin{eqnarray}
C_\ell^{(3)}&=&\frac{8}{\pi} \int dk\ k
P_\Psi(k) \Bigg[\frac{1}{3\apt_S}\left(1-\frac{1}{\apt_S(\eta_O-\eta_S)}\right)
\left(T_k(\eta_S)+\apt_S^{-1}\dot{T}_k(\eta_S)\right)j'_\ell(k(\eta_0-\eta_S))
\nonumber\\ && \qquad
 -\frac{1}{\apt_S(\eta_o-\eta_S)}\int_{\eta_S}^{\eta_0}d\eta T_k(\eta)
 j'_\ell(k(\eta_0-\eta)) + \frac{1}{\eta_0-\eta_S}
 \int_{\eta_S}^{\eta_0} d\eta\int_{\eta_S}^{\eta}d\eta' T_k(\eta')
j'_\ell(k(\eta_0-\eta'))\Bigg] \times\nonumber\\
&&\Bigg[\frac{1}{3\apt_{S'}} \left(1-
  \frac{1}{\apt_{S'}(\eta_O-\eta_{S'})} \right)
\left(T_k(\eta_{S'})+\apt_{S'}^{-1}\dot{T}_k(\eta_{S'})\right)
j'_\ell(k(\eta_0-\eta_{S'})) 
\nonumber\\ && \qquad
 -\frac{1}{\apt_{S'}(\eta_O-\eta_{S'})}\int_{\eta_{S'}}^{\eta_0}d\eta T_k(\eta)
 j'_\ell(k(\eta_0-\eta)) + \frac{1}{\eta_0-\eta_{S'}}
 \int_{\eta_{S'}}^{\eta_0} d\eta\int_{\eta_{S'}}^{\eta}d\eta' T_k(\eta')
j'_\ell(k(\eta_0-\eta'))    \Bigg]
\nonumber\\
&+&\frac{2}{\pi(\eta_O-\eta_S)} \int dk\ k P_\Psi(k)
\int_{\eta_S}^{\eta_0}d\eta\int_{\eta_S}^{\eta}
d\eta'(\eta'-\eta_S) T_k(\eta') \Big(j_\ell(k(\eta_0-\eta')) +
j''_\ell(k(\eta_0-\eta')) \Big)\times \nonumber\\ && 
\Bigg[\frac{2}{\eta_O-\eta_{S'}}\int_{\eta_{S'}}^{\eta_0}d\eta T_k(\eta)
j_\ell(k(\eta_0-\eta)) - \left(1+\frac{1}{\apt_{S'}(\eta_O-\eta_{S'})}\right)T_k(\eta_{S'})
j_\ell(k(\eta_0-\eta_{S'}))
 \Bigg] \nonumber \\
&&  \hspace{1.5 cm} +\hspace{0.5 cm} \eta_S \Leftrightarrow
\eta_{S'} ~. 
\end{eqnarray}

\begin{eqnarray}
C_\ell^{(4)}&=&-\frac{4}{\pi} \frac{1}{\eta_O-\eta_{S'}}\int dk\ k^2
P_\Psi(k)\Bigg[\frac{1}{3\apt_S}\left(1-\frac{1}{\apt_S(\eta_O-\eta_S)}\right)
\left(T_k(\eta_S) + \apt_S^{-1}\dot{T}_k(\eta_S)\right)
j'_\ell(k(\eta_0-\eta_S))    
\nonumber\\  && \qquad
-\frac{1}{\apt_S(\eta_O-\eta_S)} \int_{\eta_S}^{\eta_0}d\eta
T_k(\eta)j'_\ell(k(\eta_0-\eta)) + 
\frac{1}{\eta_0-\eta_S} \int_{\eta_S}^{\eta_0}d\eta
\int_{\eta_S}^{\eta}d\eta' T_k(\eta')
  j'_\ell(k(\eta_0-\eta'))\Bigg]\times
\nonumber\\   &&  \quad
  \int_{\eta_{S'}}^{\eta_0}d\eta \int_{\eta_{S'}}^{\eta}d\eta'(\eta'-\eta_{S'})
T_k(\eta') \Big( j_\ell(k(\eta_0-\eta')) + j''_\ell(k(\eta_0-\eta'))\Big)
 \hspace{0.5 cm} +\hspace{0.5 cm} \eta_S \Leftrightarrow \eta_{S'}~.
\end{eqnarray}

\begin{eqnarray}
C_\ell^{(5)}&=&\frac{2}{\pi}
 \frac{1}{(\eta_O-\eta_{S})(\eta_O-\eta_{S'})} \int dk\ k^3
P_\Psi(k)\times \nonumber\\
&&\Bigg[\int_{\eta_S}^{\eta_0}d\eta\int_{\eta_S}^{\eta}d\eta'(\eta'-\eta_S)
T_k(\eta')\Big(j_\ell(k(\eta_0-\eta'))+j''_\ell(k(\eta_0-\eta'))\Big)\Bigg]
 \nonumber\\ 
&\times&\Bigg[\int_{\eta_{S'}}^{\eta_0} d\eta \int_{\eta_{S'}}^{\eta}
 d\eta' (\eta'-\eta_{S'}) T_k(\eta') \Big(j_\ell(k(\eta_0-\eta')) +
 j''_\ell(k(\eta_0-\eta'))\Big) \Bigg]~. \label{eq:cl5}
\end{eqnarray}

\section{Integrals and approximations}  \label{appe3}
Here we make full use of the relatively crude approximation~(\ref{eq:I(f)})
\be
\int_{x_1}^{x_2}dx f(x)j_\ell(x) \simeq I_\ell
f(\ell)\theta(x_2-\ell)\theta(\ell-x_1)~, 
\label{eq:app}
\ee
where  $\theta$ denotes the Heaviside function, $\theta(x) = 1$ if
$x>0$ and  $\theta(x) = 0$ else. Hence we neglect
contributions to the integral which do not come from the region of
the first peak of the Bessel function. This procedure is very useful
to estimate the result, but cannot be trusted better than within a
factor of about 2. We have tested it with numerical
examples~\footnote{We must of course make sure that the above integral
converges for large $x_2$. Otherwise the approximation is not
meaningful}. A more 
detailed numerical treatment will be presented elsewhere~\cite{next}.
Furthermore, we assume a scale-invariant spectrum with $P_\Psi = A
\simeq 10^{-10}$. We also use the fact that in a matter dominated
universe the transfer function does not depend on time and can be
taken outside the time-integrals.

We define $b_S=\frac{\eta_S}{\eta_O}=\frac{1}{\sqrt{1+z_S}}$,  
 $x_S = k(\eta_0-\eta_S)$ and
 $\alpha_S=\beta\left(\frac{b_\mr{eq}}{1-b_S}\right)^4$. Note that
$x_{S'}=\frac{1-b_{S'}}{1-b_S}x_S$. 
In terms of these variables, the transfer function becomes 
\be
T^2(x_S)=\frac{1}{1+\al_S x_S^4}    ~,
\ee
 except for the $C_\ell^{(5)}$, where we have to  
take into account the log-correction.

\subsection{$C_\ell^{(1)}$}

\bea
C_\ell^{(1)}(z_S,z_{S'}) &=&
\frac{2A}{\pi(1-b_{S'})}\Bigg\{\frac{(2-b_S)(2-b_{S'})}{4(1-b_S)}
\int_0^\infty\frac{dx_S}{x_S}T^2(x_S)j_\ell(x_S)j_\ell(x_S') \nonumber\\
&+&4(1-b_S)\int_0^\infty\frac{dx_S}{x^3_S}T^2(x_S)
\left(\int_0^{x_S}dx j_\ell(x)\right)\left(\int_0^{x_{S'}}dx j_\ell(x)\right) \nonumber\\  
&-&(2-b_{S'})\int_0^\infty\frac{dx_S}{x^2_S}T^2(x_S)
\left(\int_0^{x_S}dx j_\ell(x)\right)\cdot j_\ell(x_{S'}) + b_S \Leftrightarrow b_{S'}\Bigg\} ~.
\eea

For the first term, the integral converges without the transfer
function, we may therefore neglect it and perform the integral
analytically. For the second and the third term, we use the
approximation ~(\ref{eq:app}). Assuming that $z_S<z_{S'}$  (if  not,
we reverse $z_S$ and $z_{S'}$ in the formula), we obtain   
\bea
C_\ell^{(1)}(z_S,z_{S'}) &=& \frac{2A}{\pi} \Bigg\{ 4I^2_\ell
\frac{1-b_S}{1-b_{S'}} \int_\ell^\infty
\frac{dx_S}{x^3_S}\frac{1}{1+\al_S x^4_S} 
-\frac{I^2_\ell}{\ell^2}\frac{2-b_S}{1-b_{S'}}\frac{1}{1+\ell^4\alpha_S}
\nonumber\\ 
&+& \frac{\sqrt\pi}{16}\frac{\Gamma(\ell)}{\Gamma(\ell+3/2)}(2-b_S)(2-b_{S'}) 
\frac{(1-b_S)^{\ell-1}(1-b_{S'})^{\ell-1}}{(2-b_S-b_{S'})^{2\ell}}
F\left(\ell,\ell+1;2\ell+2;\frac{4(1-b_S)(1-b_{S'})}{(2-b_S-b_{S'})^2}
\right) \Bigg\} ~.\nonumber\\ 
\eea
Here $F$ denotes the hyper-geometric function and $\Ga$ is the
$\Ga$--function. We use the notation and normalization of~\cite{AS}.

\subsection{$C_\ell^{(2)}$}

\bea
C_\ell^{(2)}(z_S,z_{S'})&=&
\frac{-2A}{\pi(1-b_{S'})}\Bigg\{-\frac{(2-3b_S)(2-b_{S'})}{2(1-b_S)}
\int_0^\infty \frac{dx_S}{x_S} T^2(x_S)j_\ell(x_S)j_\ell(x_S') \nonumber\\
&-&\frac{b_S(2-3b_S)(2-b_{S'})}{12(1-b_S)^2}
\int_0^\infty dx_S T^2(x_S)j'_l(x_S)j_\ell(x_{S'}) \nonumber\\
&-&4(1-b_S)\int_0^\infty\frac{dx_S}{x^3_S}T^2(x_S)
\left(\int_0^{x_S}dx j_\ell(x)\right)\left(\int_0^{x_{S'}}dx j_\ell(x)\right) \nonumber\\  
&+&(6-7b_{S'})\int_0^\infty\frac{dx_S}{x^2_S}T^2(x_S)
\left(\int_0^{x_S}dx j_\ell(x)\right)\cdot j_\ell(x_{S'})\nonumber\\
&+&\frac{b_{S'}(2-3b_{S'})}{3(1-b_S)}\int_0^\infty\frac{dx_S}{x_S}T^2(x_S)
\left(\int_0^{x_S}dx j_\ell(x)\right)\cdot j'_l(x_{S'})\Bigg\} + b_S \Leftrightarrow b_{S'} ~.
\eea

Here again, the terms which contain only an integral over $x_S$ can be
calculated analytically when we neglect the decay of the transfer
function. For the other terms we use the approximation ~(\ref{eq:app}).

\bea
C_\ell^{(2)}(z_S,z_{S'}) &=& 
\frac{-2A}{\pi(1-b_S)(1-b_{S'})}\Bigg\{-8I^2_\ell(1-b_S)^2\int_\ell^\infty 
\frac{dx_S}{x^3_S}\frac{1}{1+\al_S x^4_S}\nonumber\\ 
&+&\frac{I^2_\ell}{\ell^2}\frac{1}{1+\alpha_S\ell^4}\Big[(6-7b_S)(1-b_S)
+\frac{4b_S(2-3b_S)}{3}\frac{\alpha_S\ell^4}{1+\alpha_S\ell^4}\Big] \nonumber\\
&-& \frac{I_\ell}{\ell^2(2\ell+1)} \frac{b_S(2-3b_S)}{3}
\frac{1}{1+\alpha_S\ell^4} 
\Big[\ell(\ell-1)I_{\ell-1}\theta\left(\ell-(\ell-1)\frac{1-b_S}{1-b_{S'}}
  \right) \nonumber \\ && \qquad \qquad
-(\ell+1)^2I_{\ell+1}\theta\left(\ell-(\ell+1)\frac{1-b_S}{1-b_{S'}}\right)
\Big]\nonumber\\ 
&-& \frac{I_\ell}{\ell^2(2\ell+1)} \frac{b_{S'}(2-3b_{S'})}{3}
\frac{1}{1+\alpha_{S'}\ell^4} 
\Big[\ell(\ell-1)I_{\ell-1}\theta\left(\ell-(\ell-1)\frac{1-b_{S'}}{1-b_{S}}
  \right)  \nonumber \\ && \qquad \qquad
-(\ell+1)^2I_{\ell+1}\theta\left(\ell-(\ell+1)\frac{1-b_{S'}}{1-b_{S}}\right)
\Big]\nonumber\\ 
&-&\big(4(1-b_S-b_{S'})+3b_Sb_{S'})\frac{\sqrt\pi}{4}\frac{\Gamma(\ell)}{\Gamma(\ell+3/2)} 
\frac{(1-b_S)^{\ell}(1-b_{S'})^{\ell}}{(2-b_S-b_{S'})^{2\ell}}
F\left(\ell,\ell+1;2\ell+2;\frac{4(1-b_S)(1-b_{S'})}{(2-b_S-b_{S'})^2} \right) \nonumber\\
&-&\frac{b_S(2-3b_S)(2-b_{S'})}{12(2\ell+1)}\Big[\frac{\sqrt\pi}{2}\frac{\Gamma(\ell+1)}{\Gamma(\ell+1/2)} 
\frac{(1-b_S)^{\ell-1}}{(1-b_{S'})^{\ell}}
F\left(\ell,-1/2;\ell+1/2;\frac{(1-b_S)^2}{(1-b_{S'})^2} \right) \nonumber\\
&&-\frac{\sqrt\pi}{4}\frac{\Gamma(\ell+2)}{\Gamma(\ell+5/2)} 
\frac{(1-b_S)^{\ell+1}}{(1-b_{S'})^{\ell+2}}
F\left(\ell+1,1/2;\ell+5/2;\frac{(1-b_S)^2}{(1-b_{S'})^2} \right)\Big]\nonumber\\
&-&\frac{b_{S'}(2-3b_{S'})(2-b_{S})}{12(2\ell+1)}\Big[\frac{\sqrt\pi}{4}\frac{\Gamma(\ell+1)}{\Gamma(\ell+3/2)} 
\frac{(1-b_S)^{\ell}}{(1-b_{S'})^{\ell+1}}
F\left(\ell,1/2;\ell+3/2;\frac{(1-b_S)^2}{(1-b_{S'})^2} \right) \nonumber\\
&&-\frac{\sqrt\pi}{2}\frac{\Gamma(\ell+2)}{\Gamma(\ell+3/2)} 
\frac{(1-b_S)^{\ell}}{(1-b_{S'})^{\ell+1}}
F\left(\ell+1,-1/2;\ell+3/2;\frac{(1-b_S)^2}{(1-b_{S'})^2} \right)\Big]
\Bigg\} ~.
\eea

\subsection{$C_\ell^{(3)}$}

\bea
C_\ell^{(3)}(z_S,z_{S'})&=&
\frac{2A}{\pi(1-b_S)(1-b_{S'})}\Bigg\{\Big(2-\frac{9(b_S+b_{S'})}{2}+8b_Sb_{S'}\Big)
\int_0^\infty \frac{dx_S}{x_S} T^2(x_S)j_\ell(x_S)j_\ell(x_{S'}) \nonumber\\
&+&\frac{b_S(2-3b_S)(2-3b_{S'})}{6(1-b_S)}
\int_0^\infty dx_S T^2(x_S)j'_l(x_S)j_\ell(x_{S'}) + b_S \Leftrightarrow b_{S'} \nonumber\\
&+&\frac{b_Sb_{S'}(2-3b_S)(2-3b_{S'})}{36(1-b_S)^2}
\int_0^\infty dx_S x_S T^2(x_S)j'_l(x_S)j'_l(x_{S'})\nonumber\\
&-&4(1-b_S)^2\int_0^\infty\frac{dx_S}{x^3_S}T^2(x_S)
\left(\int_0^{x_S}dx j_\ell(x)\right)\left(\int_0^{x_{S'}}dx j_\ell(x)\right) \nonumber\\  
&+&3b_{S'}(1-b_S)\int_0^\infty\frac{dx_S}{x^2_S}T^2(x_S)
\left(\int_0^{x_S}dx j_\ell(x)\right)\cdot j_\ell(x_{S'})+ b_S \Leftrightarrow b_{S'}\nonumber\\
&-&\frac{b_{S'}(2-3b_{S'})}{3}\int_0^\infty\frac{dx_S}{x_S}T^2(x_S)
\left(\int_0^{x_S}dx j_\ell(x)\right)\cdot j'_l(x_{S'})+ b_S \Leftrightarrow b_{S'}\nonumber\\
&+&2(1-b_S)^2\int_0^\infty\frac{dx_S}{x^3_S}T^2(x_S)
\left(\int_0^{x_S}dx \int_x^{x_S}dx'(x_S-x') j_\ell(x')\right)\cdot \left(\int_0^{x_{S'}}dx j_\ell(x)\right)
+ b_S \Leftrightarrow b_{S'}\nonumber\\
&-&\frac{(2-b_{S'})(1-b_S)}{2}\int_0^\infty\frac{dx_S}{x^2_S}T^2(x_S)
\left(\int_0^{x_S}dx \int_x^{x_S}dx'(x_S-x') j_\ell(x')\right)\cdot j_\ell(x_{S'})
+ b_S \Leftrightarrow b_{S'} \Bigg\} ~.
\eea

Here, it is not possible to neglect the transfer function in the third
integral, because for $z_S=z_{S'}$ the integral does not converge
without $T^2(x_S)$. We therefore  have to calculate the third (the
Doppler term) term numerically. 
 
\bea
C_\ell^{(3)}(z_S,z_{S'}) &=&\frac{2A}{\pi(1-b_S)(1-b_{S'})}\Bigg\{-2I^2_\ell (1-b_S)^2\int_\ell^\infty
\frac{dx_S}{x^3_S}\frac{1}{1+\al_S x^4_S}\Big(2(1+\ell^2)+\frac{\ell x_S}{1-b_S}(b_S+b_{S'}-2)\Big)\nonumber\\ 
&+&\frac{I^2_\ell}{\ell^2}\frac{1}{1+\alpha_S\ell^4}\Big[3b_S(1-b_S)
-\ell^2(1-\frac{b_S}{2})(b_S-b_{S'})+b_S(b_S-2/3)
\frac{4\alpha_S\ell^4}{1+\alpha_S\ell^4}\Big] \nonumber\\ 
&+& \frac{I_\ell}{\ell^2(2\ell+1)}\frac{b_S(2-3b_S)}{3}
\frac{1}{1+\alpha_S\ell^4} 
\Big[\ell(\ell-1)I_{\ell-1}\theta\left(\ell-(\ell-1)
  \frac{1-b_S}{1-b_{S'}}\right)    \nonumber \\ && \qquad \qquad
-(\ell+1)^2I_{\ell+1}\theta\left(\ell-(\ell+1)
   \frac{1-b_S}{1-b_{S'}}\right) \Big]\nonumber\\
&+& \frac{I_\ell}{\ell^2(2\ell+1)}\frac{b_{S'}(2-3b_{S'})}{3}\frac{1}{1+\alpha_{S'}\ell^4}
\Big[\ell(\ell-1)I_{\ell-1}\theta\left(\ell-(\ell-1)
\frac{1-b_{S'}}{1-b_{S}}\right)  \nonumber \\ && \qquad \qquad
-(\ell+1)^2I_{\ell+1}\theta\left(\ell-(\ell+1)\frac{1-b_{S'}}{1-b_{S}}
\right) \Big]\nonumber\\
&+&\big(2-\frac{9}{2}(b_S+b_{S'})+8b_Sb_{S'})\frac{\sqrt\pi}{4}
\frac{\Gamma(\ell)}{\Gamma(\ell+3/2)}  
\frac{(1-b_S)^{\ell}(1-b_{S'})^{\ell}}{(2-b_S-b_{S'})^{2\ell}}
F\left(\ell,\ell+1;2\ell+2;\frac{4(1-b_S)(1-b_{S'})}{(2-b_S-b_{S'})^2} \right) \nonumber\\
&+&\frac{b_S(2-3b_S)(2-3b_{S'})}{6(2\ell+1)}\Big[\frac{\sqrt\pi}{2}\frac{\Gamma(\ell+1)}{\Gamma(\ell+1/2)} 
\frac{(1-b_S)^{\ell-1}}{(1-b_{S'})^{\ell}}
F\left(\ell,-1/2;\ell+1/2;\frac{(1-b_S)^2}{(1-b_{S'})^2} \right) \nonumber\\
&&-\frac{\sqrt\pi}{4}\frac{\Gamma(\ell+2)}{\Gamma(\ell+5/2)} 
\frac{(1-b_S)^{\ell+1}}{(1-b_{S'})^{\ell+2}}
F\left(\ell+1,1/2;\ell+5/2;\frac{(1-b_S)^2}{(1-b_{S'})^2} \right)\Big]\nonumber\\
&+&\frac{b_{S'}(2-3b_{S'})(2-3b_{S})}{6(2\ell+1)}\Big[\frac{\sqrt\pi}{4}\frac{\Gamma(\ell+1)}{\Gamma(\ell+3/2)} 
\frac{(1-b_S)^{\ell}}{(1-b_{S'})^{\ell+1}}
F\left(\ell,1/2;\ell+3/2;\frac{(1-b_S)^2}{(1-b_{S'})^2} \right) \nonumber\\
&&-\frac{\sqrt\pi}{2}\frac{\Gamma(\ell+2)}{\Gamma(\ell+3/2)} 
\frac{(1-b_S)^{\ell}}{(1-b_{S'})^{\ell+1}}
F\left(\ell+1,-1/2;\ell+3/2;\frac{(1-b_S)^2}{(1-b_{S'})^2} \right)\Big]\nonumber\\
&+&\frac{b_Sb_{S'}(2-3b_S)(2-3b_{S'})}{36(1-b_S)^2}\int_0^\infty dx_S x_S T^2(x_S)j'_\ell(x_S)j'_\ell(x_{S'})
\Bigg\}  ~.
\eea
The last term in this sum is determined by numerical integration over $x_S$.
\subsection{$C_\ell^{(4)}$}

\bea
C_\ell^{(4)}(z_S,z_{S'})&=&
\frac{-2A}{\pi(1-b_S)(1-b_{S'})}\Bigg\{(2-3b_S)(1-b_{S'})
\int_0^\infty \frac{dx_S}{x_S} T^2(x_S)j_\ell(x_S)j_\ell(x_{S'}) \nonumber\\
&+&\frac{b_S(2-3b_S)(1-b_{S'})}{6(1-b_S)}
\int_0^\infty dx_S T^2(x_S)j'_l(x_S)j_\ell(x_{S'}) \nonumber\\
&+&4(1-b_S)^2\int_0^\infty\frac{dx_S}{x^3_S}T^2(x_S)
\left(\int_0^{x_S}dx j_\ell(x)\right)\left(\int_0^{x_{S'}}dx j_\ell(x)\right) \nonumber\\  
&-&2(3-4b_{S'})(1-b_S)\int_0^\infty\frac{dx_S}{x^2_S}T^2(x_S)
\left(\int_0^{x_S}dx j_\ell(x)\right)\cdot j_\ell(x_{S'})\nonumber\\
&-&\frac{b_{S'}(2-3b_{S'})}{3}\int_0^\infty\frac{dx_S}{x_S}T^2(x_S)
\left(\int_0^{x_S}dx j_\ell(x)\right)\cdot j'_l(x_{S'})\nonumber\\
&-&2(1-b_S)^2\int_0^\infty\frac{dx_S}{x^3_S}T^2(x_S)
\left(\int_0^{x_{S}}dx j_\ell(x)\right)\cdot\left(\int_0^{x_{S'}}dx \int_x^{x_{S'}}dx'(x_{S'}-x') j_\ell(x')\right) 
\nonumber\\ 
&+&(2-3b_{S})(1-b_{S})\int_0^\infty\frac{dx_S}{x^2_S}T^2(x_S)
\left(\int_0^{x_{S'}}dx \int_x^{x_{S'}}dx'(x_{S'}-x') j_\ell(x')\right)\cdot j_\ell(x_{S})\nonumber\\
&+&\frac{b_S(2-3b_{S})}{6}\int_0^\infty\frac{dx_S}{x_S}T^2(x_S)
\left(\int_0^{x_{S'}}dx \int_x^{x_{S'}}dx'(x_{S'}-x') j_\ell(x')\right)\cdot j'_l(x_{S})\Bigg\}\nonumber\\
&+& b_S \Leftrightarrow b_{S'}  ~.
\eea

\bea
C_\ell^{(4)}(z_S,z_{S'}) &=&\frac{-2A}{\pi(1-b_S)(1-b_{S'})}
\Bigg\{2I^2_\ell (1-b_S)^2\int_\ell^\infty 
\frac{dx_S}{x^3_S}\frac{1}{1+\al_S x^4_S}\Big(2(2+\ell^2) + \frac{\ell
  x_S}{1-b_S}(b_S+b_{S'}-2)\Big)\nonumber\\  
&+&\frac{I^2_\ell}{\ell^2}\frac{1}{1+\alpha_S\ell^4}\Big[-2(1-b_S)(3-4b_S)
+\ell^2(2-3b_S)(b_S-b_{S'})+b_S(b_S-2/3)
\frac{4\alpha_S\ell^4}{1+\alpha_S\ell^4}\Big] \nonumber\\ 
&+& \frac{I_\ell}{\ell^2(2\ell+1)} \frac{b_S(2-3b_S)}{3}
\frac{1}{1+\alpha_S\ell^4} \Big[ \ell(\ell-1)I_{\ell-1}
  \theta\left(\ell-(\ell-1)\frac{1-b_S}{1-b_{S'}} \right)
  \nonumber  \\  && \qquad \qquad 
-(\ell+1)^2I_{\ell+1}\theta\left(\ell-(\ell+1)\frac{1-b_S}{1-b_{S'}}\right)
\Big]\nonumber\\ 
&+& \frac{I_\ell}{\ell^2(2\ell+1)}\frac{b_{S'}(2-3b_{S'})}{3}
\frac{1}{1+\alpha_{S'}\ell^4} 
\Big[\ell(\ell-1)I_{\ell-1}\theta\left(\ell-(\ell-1)\frac{1-b_{S'}}{1-b_{S}}
  \right)    \nonumber  \\  && \qquad \qquad 
-(\ell+1)^2I_{\ell+1}\theta\left(\ell-(\ell+1)\frac{1-b_{S'}}{1-b_{S}}\right)
\Big]\nonumber\\ 
&+& \frac{I_\ell \ell}{2\ell+1}\frac{b_S(1-b_S)^3(2-3b_S)}{6}
\Big[\frac{\ell I_{\ell-1}}{\ell-1}\frac{b_{S'}-1+\ell(b_S-b_{S'})}
  {(1-b_S)^4+ (\ell-1)^4 b^4_\mr{eq}}  
\theta\left(\ell-1-\ell\frac{1-b_S}{1-b_{S'}}\right)\nonumber\\
&&-I_{\ell+1}\frac{1-b_{S'} + \ell(b_S-b_{S'}) }{(1-b_S)^4+ (\ell+1)^4
  b^4_\mr{eq}} \theta\left(\ell+1-\ell\frac{1-b_S}{1-b_{S'}}\right)\Big]+b_S
\Leftrightarrow b_{S'}\nonumber\\ 
&+&\big(4-5(b_S+b_{S'})+6b_Sb_{S'})\frac{\sqrt\pi}{4}
\frac{\Gamma(\ell)}{\Gamma(\ell+3/2)}  
\frac{(1-b_S)^{\ell}(1-b_{S'})^{\ell}}{(2-b_S-b_{S'})^{2\ell}}
F\left(\ell,\ell+1;2\ell+2;\frac{4(1-b_S)(1-b_{S'})}{(2-b_S-b_{S'})^2}
\right) \nonumber\\ 
&+&\frac{b_S(2-3b_S)(1-b_{S'})}{6(2\ell+1)}\Big[\frac{\sqrt\pi}{2}
  \frac{\Gamma(\ell+1)}{\Gamma(\ell+1/2)}   
\frac{(1-b_S)^{\ell-1}}{(1-b_{S'})^{\ell}}
F\left(\ell,-1/2;\ell+1/2;\frac{(1-b_S)^2}{(1-b_{S'})^2} \right) \nonumber\\
&&-\frac{\sqrt\pi}{4}\frac{\Gamma(\ell+2)}{\Gamma(\ell+5/2)} 
\frac{(1-b_S)^{\ell+1}}{(1-b_{S'})^{\ell+2}}
F\left(\ell+1,1/2;\ell+5/2;\frac{(1-b_S)^2}{(1-b_{S'})^2} \right)\Big]
\nonumber\\
&+&\frac{b_{S'}(2-3b_{S'})(1-b_{S})}{6(2\ell+1)}\Big[\frac{\sqrt\pi}{4}
  \frac{\Gamma(\ell+1)}{\Gamma(\ell+3/2)} 
\frac{(1-b_S)^{\ell}}{(1-b_{S'})^{\ell+1}}
F\left(\ell,1/2;\ell+3/2;\frac{(1-b_S)^2}{(1-b_{S'})^2} \right) \nonumber\\
&&-\frac{\sqrt\pi}{2}\frac{\Gamma(\ell+2)}{\Gamma(\ell+3/2)} 
\frac{(1-b_S)^{\ell}}{(1-b_{S'})^{\ell+1}}
F\left(\ell+1,-1/2;\ell+3/2;\frac{(1-b_S)^2}{(1-b_{S'})^2} \right)\Big] \Bigg\}  ~.
\eea

\subsection{$C_\ell^{(5)}$}

\bea
C_\ell^{(5)}(z_S,z_{S'})&=&
\frac{2A}{\pi(1-b_S)(1-b_{S'})}\Bigg\{(1-b_S)(1-b_{S'})
\int_0^\infty \frac{dx_S}{x_S} T^2(x_S)j_\ell(x_S)j_\ell(x_{S'}) \nonumber\\
&+&4(1-b_S)^2\int_0^\infty\frac{dx_S}{x^3_S}T^2(x_S)
\left(\int_0^{x_S}dx j_\ell(x)\right)\left(\int_0^{x_{S'}}dx
j_\ell(x)\right) \nonumber\\   
&-&2(1-b_S)^2\int_0^\infty\frac{dx_S}{x^2_S}T^2(x_S)
\left(\int_0^{x_{S'}}dx j_\ell(x)\right)\cdot j_\ell(x_S) + b_S
\Leftrightarrow b_{S'}\nonumber\\ 
&-&2(1-b_S)^2\int_0^\infty\frac{dx_S}{x^3_S}T^2(x_S)
\left(\int_0^{x_{S'}}dx j_\ell(x)\right)\left(\int_0^{x_{S}}dx
\int_x^{x_{S}}dx'(x_{S}-x') j_\ell(x')\right)  
+ b_S \Leftrightarrow b_{S'}\nonumber\\ 
&+&(1-b_{S})(1-b_{S'})\int_0^\infty\frac{dx_S}{x^2_S}T^2(x_S)
\left(\int_0^{x_S}dx \int_x^{x_{S}}dx'(x_{S}-x')
j_\ell(x')\right) j_\ell(x_{S'})+ b_S \Leftrightarrow b_{S'} \nonumber\\ 
&+&(1-b_{S})^2\int_0^\infty\frac{dx_S}{x^3_S}T^2(x_S)
\left(\int_0^{x_{S}}dx \int_x^{x_{S}}dx'(x_{S}-x') j_\ell(x')\right)
\left(\int_0^{x_{S'}}dx \int_x^{x_{S'}}dx'(x_{S'}-x')
j_\ell(x')\right)\Bigg\}~.\nonumber\\
\eea

The first term is dominated on large scale and we may thus set $T\equiv
1$ so that it can be integrated analytically. For the other terms we
use again the approximation 
(\ref{eq:app}) for the integrals $dx$ or $dx'$. The biggest
contribution then comes from the last term where we have to perform
two double integrals $dxdx'$, which result in $I_\ell^2(2+\ell^2-\ell
x_S)(2+\ell^2-\ell x_{S'}) \propto \ell^3$. In this term, which becomes
large for large $\ell$ or large $x_S$, we take into account the
log-correction to the transfer function for better accuracy. From the
expression in Ref.~\cite{Dod} and our definitions we find: 

\be
T^2(x_S)=\frac{1}{1+\frac{\al_S x^4_S}
{\ln^2\left(1+\frac{7.8\cdot 10^{-4}}{1-b_S}x_S \right)}} ~.
\ee
 
Using our approximation ~(\ref{eq:app}), we obtain

\bea
C_\ell^{(5)}(z_S,z_{S'}) &=&\frac{2A}{\pi(1-b_S)(1-b_{S'})}\Bigg\{
I^2_\ell (1-b_S)^2\int_\ell^\infty 
\frac{dx_S}{x^3_S}\frac{1}{1+\frac{\al_S x^4_S}{\ln^2\left(1 +
    \frac{7.8\cdot 10^{-4}}{1-b_S}x_S \right)}} 
(2+\ell^2-\ell x_S)(2+\ell^2-\ell x_{S'})\nonumber\\ 
&-&\frac{I^2_\ell}{\ell^2}\frac{1-b_S}{1+\hat{\alpha}_S\ell^4}
\Big(2-(2+\ell^2)b_S + \ell^2b_{S'}\Big) 
\nonumber\\
&+&\frac{\sqrt\pi}{4}\frac{\Gamma(\ell)}{\Gamma(\ell+3/2)} 
\frac{(1-b_S)^{\ell+1}(1-b_{S'})^{\ell+1}}{(2-b_S-b_{S'})^{2\ell}}
F\left(\ell,\ell+1;2\ell+2;\frac{4(1-b_S)(1-b_{S'})}{(2-b_S-b_{S'})^2}
\right) \Bigg\} ~,
\eea

where 
\be
\hat{\alpha}_S=\beta\left(\frac{b_\mr{eq}}{1-b_S}\right)^4
\frac{1}{\ln^2\left(1+\frac{7.8\cdot 10^{-4}\ell}{1-b_S}\right)}~.
\ee
The remaining integral represents by far the largest contribution to
$C_\ell^{(5)}$. For sources with equal redshifts $z_S=z_{S'}=z$, the spectrum
$C_\ell^{(5)}(z,z)$ grows 
until $\hat\al_s \ell^4 \sim 1$ and decays for larger
$\ell$. Neglecting the log correction we have $\al_S =
\left(\beta^{1/4}\frac{b_\mr{eq}}{1-b_S}\right)^4 \equiv
\ell_{\max}^{-4}$.
Hence $C_\ell^{(5)}$ grows roughly until $\ell_{\max}$ and decays afterwards.
With $b_\mr{eq} = (\eta_\mr{eq}/\eta_O)\simeq 0.01$ we obtain
$$
\ell_{\max} \simeq 760 \frac{\sqrt{z_S+1}-1}{\sqrt{1+z_S}}~.
$$

For a crude order of magnitude estimate, we first neglect the log
corrections. For $\ell\ll \ell_{\max}$ the integral is dominated by
the region $x_S< \ell_{\max}$ and we may simply integrate until
$x_S \simeq \ell_{\max}$, neglecting the $x_S^4$ decay of the
transfer function. In the opposite region, if $\ell \gg \ell_{\max}$, we
may neglect the $1$ in the denominator of the integral. An interpolation
between this two asymptotic regimes gives
\be\label{eq:C5app}
C_\ell^{(5)}(z_S,z_{S}) \simeq \frac{2AI_\ell^2\ell^2}{\pi}\left\{ 
\begin{array}{ll} \ln\left(\frac{\ell_{\max}}{\ell}\right) + \frac{1}{4} &
  \mbox{ if } \quad \ell < \ell_{\max} \\
\frac{1}{4}\left(\frac{\ell_{\max}}{\ell}\right)^4 &
 \mbox{ if } \quad \ell > \ell_{\max} ~. \end{array} \right.
\ee  
Since $I_\ell^2 \propto 1/\ell$ we see that $\ell(\ell+1)C_\ell^{(5)}$
grows like $\ell^3$ for small $\ell$'s and it decays like $1/\ell$ for
large $\ell$'s. The broad maximum is reached roughly at $\ell_{\max}
\simeq 760 \frac{\sqrt{1+z_S}-1}{\sqrt{z_S+1}} =760(1-b_S)$ and is of
the order of 
$(A/\pi)\ell_{\max}^3$.  This approximation is, however, surprisingly
bad. We therefore take into account the log in the transfer function by
simple replacing $\al_S$ by $\hat\al_S$, where $\ell$ in the
expression for $\hat\al_S$ denotes the lower boundary of the integral. The
expression for $\ell_{\max}$ then becomes $\ell$--dependent,
\be \label{eq:lmaxl}
\ell_{\max} \simeq
\frac{\sqrt{\ln(1+7.8\times 10^{-4}\ell/(1-b_{S}))}}
                                     {\beta^{1/4}b_\mr{eq}} (1-b_S) ~.
\ee
For $\ell <1.3\times 10^3 (1-b_S) \equiv \ell_S$ the log can be expanded and
$\ell_{\max}/\ell$ behaves like $\ell^{-1/2}$  leading to a linear growth of
$\ell(\ell+1)C_\ell^{(5)}$. Only above $\ell_S$ it levels off. For
$z_S=2$, the asymptotic regime, where $\ell(\ell+1)C_\ell^{(5)}$ decays like
$1/\ell$ is actually only reached at $\ell \sim 2000$, where our
approximations (and linear perturbation theory) no longer hold.

\begin{figure}[ht]
\centerline{\epsfig{figure=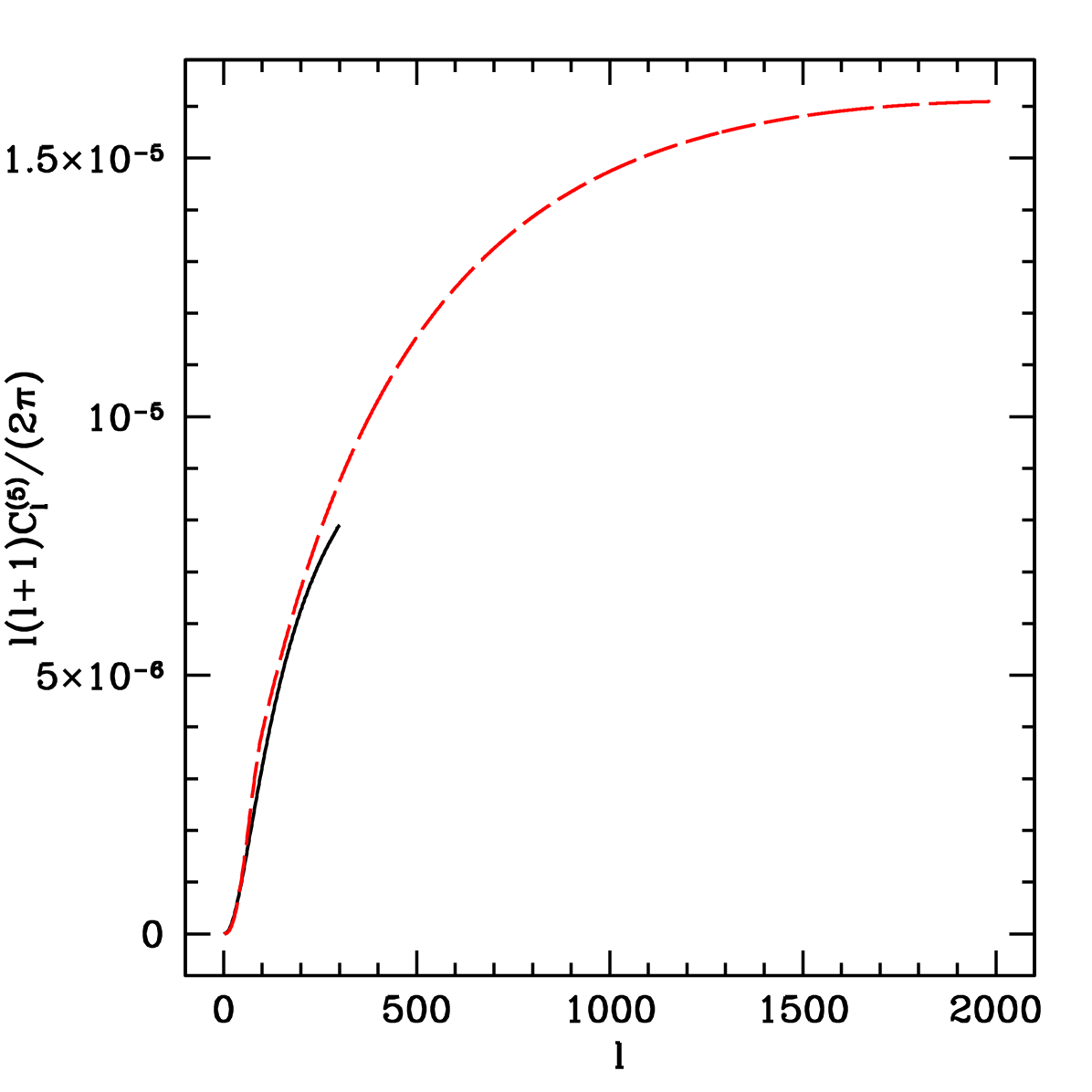,width=6.5cm}}
\caption{ \label{fig:C5app} The approximation for
  $\ell(\ell+1)C_\ell^{(5)}(z,z')/(2\pi)$ given in
  Eq.~(\protect\ref{eq:C5app}) (red, dashed line) is compared with our
  numerical result (black, solid line) for $z=2$.
}
\end{figure}

In Fig.~\ref{fig:C5app} we plot the approximation given in
Eq.~(\ref{eq:C5app}) with $\ell_{\max}$ given
in~(\ref{eq:lmaxl}) for $z_S=z_{S'}=2$ and hence $\ell_S\simeq
540$. Actually, to have a better fit with the numerical 
integral we choose a slightly modified value, namely $\tilde \ell_{\max} 
 = 0.75\ell_{\max}$.

\twocolumngrid

\end{document}